\documentclass[journal = cgdefu, manuscript=article, dvips]{achemso}
\usepackage{graphicx}
\usepackage{graphics}
\usepackage{epstopdf}
\usepackage{latexsym}
\usepackage{amssymb,amsmath}
\usepackage{enumitem}

\usepackage{multirow}

\usepackage{color}

\def\vc#1{{\bf #1}}

\makeatletter
\def\@seccntformat#1{\csname the#1\endcsname.\quad}
\makeatother

\renewcommand{\theenumi}{\roman{enumi}}

\title[Growth rate of crystal surface]
{Growth rate of crystal surfaces  
with several dislocation centers}

\author{Takeshi Ohtsuka}
\email{tohtsuka@gunma-u.ac.jp}
\affiliation{Division of Pure and Applied Science,
Faculty of Science and Technology, Gunma University,
4-2, Aramaki-machi, Maebashi, Gunma 371-8510, Japan}

\author{Yen-Hsi Richard Tsai}
\affiliation{Department of Mathematics, Institute for
Computational Engineering and Science(ICES),
The University of Texas at Austin, Austin, TX78712, USA}
\alsoaffiliation{KTH Royal Institute of Technology, Stockholm, Sweden}

\author{Yoshikazu Giga}
\affiliation{Graduate School of Mathematical Sciences,
The University of Tokyo,
Komaba 3-8-1, Meguro-ku, Tokyo 153-8914, Japan}

\begin{document}

\maketitle

\begin{abstract}
 We study analytically and numerically
 the growth rate
 of a crystal surface growing by several screw dislocations.
 To describe several spiral steps we use
 the revised level set method for spirals by the authors
 (Journal of Scientific Computing 62, 831-874, 2015).
 We carefully compare our simulation results
 on the growth rates with predictions
 in a classical paper by Burton et al.
 (Philos Trans R Soc Lond Ser A Math Phys Sci 243,299-358, 1951).
 Some discrepancy between the growth rate
 computed by our method and reported by the classical paper
 is observed.
 In this paper we propose improved estimates on the growth rate
 with several different configurations.
 In particular we give a quantitive definition of
 the critical distance of co-rotating screw dislocations
 under which the effective growth resembles that of a single spiral.
 The proposed estimates are in agreement
 with our numerical simulations.
 The influence of distribution of screw dislocations
 in a group on a line to the growth rate,
 and the growth rate by a group including different
 rotational orientations of spirals are also studied in this paper.
\end{abstract}



\section{Introduction}
\label{intro}
We are interested in modeling and simulation of growth of crystal surfaces that have discontinuities in height  along curves that spiral out from a few centers. The centers correspond physically to the end points of screw dislocation in the crystalline structure.  
Due to the dislocations, the crystal surface have discontinuities which are generally referred to as steps. 
Spiral steps evolve by catching atoms
on the surface, and the increase in crystal height could be thought of as 
the spiral steps climbing up the helical surface provided by
lattice structure of atoms including 
screw dislocations.
We refer such type of crystal growth as
``screw dislocation aided crystal growth''.
\begin{figure}[htbp]
 \begin{center}
  \includegraphics[scale=1.0]{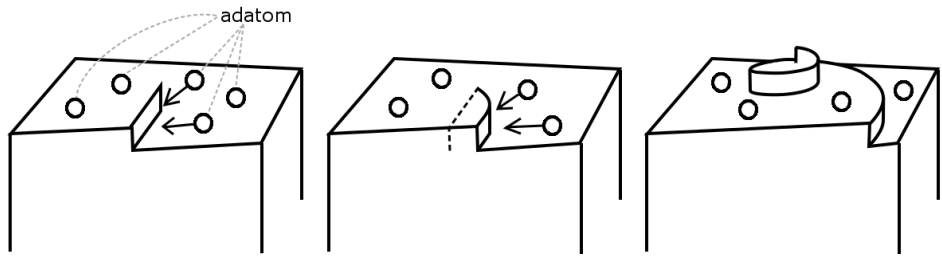}
 \end{center}
 \caption{Illustration of crystal growth with aid of screw dislocation.}
\end{figure}

Since the spiral dynamics of several screw dislocations
involve merging of different spirals,
implicit interface methods are attractive options for description of the spiral steps.
There are several nice books on
details of the convenntional level set method\cite{Osher-Sethian-level-set:1988, sethian-book:1999, Osher-Fedkiw:2000}
and of its mathematical foundation \cite{Giga:2006}.
For spiral curves,
the authors introduce 
the level set formulation 
\cite{Ohtsuka:2003wi, OTG:2015JSC}.
On the other hand, several phase-field approaches for evolving spirals
are
introduced \cite{Karma:1998, Kobayashi:1990th, Miura:2015CGD}.

In this paper, we study the growth rates of such crystals as described
in the classical paper by Burton et al. \cite{Burton:1951tr}
using the method proposed by the authors \cite{OTG:2015JSC}.
In particular, we give a quantitative definition of the critical distance
(of co-rotating screw dislocations)
under which the effective growth resembles that of a single spiral.
We conclude that the critical distance
predicted by Burton et al. \cite{Burton:1951tr} is
too small compared with our definition.
We further give some improved estimates
of the growth rate of crystal surface
by co-rotating spirals.
Finally, we present a numerical study on
growth rates by a group of screw dislocations.
In particular, the influence of distribution screw dislocations
in a group of them is considered.
Recently, Miura--Kobayashi \cite{Miura:2015CGD} proposed a
phase-filed formulation for spiral crystal growth,
and they concluded that their numerical simulations
agree with the prediction of Burton et al \cite{Burton:1951tr}.
One of the aims of this paper is to clarify some discrepancy
between the growth rates computed by our method and those
reported in previous papers \cite{Burton:1951tr, Miura:2015CGD}.
Moreover, we study on the growth rate by a group
including several rotational orientation,
which is mentioned
by Burton et al. \cite{Burton:1951tr}
but not treated by Miura--Kobayashi \cite{Miura:2015CGD}.

The numerical simulations reported in this paper were computed
by an implementation of the algorithm
proposed by the authors \cite{OTG:2015JSC},
and it is reviewed in the next section.

\section{Preliminaries}
\label{preliminaries}

In this section we recall
the level set method \cite{Ohtsuka:2003wi, OTG:2015JSC}
for evolving spiral steps by \eqref{step velocity}
on the crystal surface.
The method also includes a way to reconstruct the crystal surface
from the solution of the level set equation.

We consider a growing crystal surface that contains spiral steps attached to many screw dislocations.
These steps are modelled as curves in $\mathbb{R}^2$ in this paper, and we will use ``curves'' or ``steps''
interchangeably in this paper.
According to the theory of Burton et al. \cite{Burton:1951tr},
spiral steps move with normal velocties 
given as
\begin{equation}
 \label{step velocity}
  V = v_\infty (1 - \rho_c \kappa),
\end{equation}
where $\kappa$ is the curvature corresponding
to the inverse direction of the evolution of steps,
$v_\infty$ and $\rho_c$ are positive constants describing
the velocity of straight line steps and the critical radius
of the two dimensional kernel, respectively.

When a single spiral step with a height, $h_0 > 0$,
steadily rotates with angular velocity $\omega$,
then the surface grows with the vertical growth rate
\[
 R = \frac{\omega h_0}{2 \pi}.
\] 
Burton et al. \cite{Burton:1951tr}  calculated $\omega$
by approximating the form of the spiral step
with an Archimedean spiral, and then
they obtained that $\omega = v_\infty / (2 \rho_c)$.

Our focus is naturally on the growth rate of crystals
that evolve under the presence of many steps.
Some heuristic observation
on such settings was discussed
in the classical paper by Burton et al.\cite{Burton:1951tr} 
However, it was pointed out that the estimate
on the growth rate for such setups was not accurate.

\subsection{Description of spirals}

Let  $\Omega$ be a bounded region in $\mathbb{R}^2$,
and $\vc{a}_1, \vc{a}_2, \ldots, \vc{a}_N \in \Omega$ be the centers of the spirals.
Define 
\[
W = \Omega \setminus \bigcup_{j=1}^N B_r(\vc{a}_j), 
\]
where $B_r(\vc{a}_j)$ is a closed disc with radius $r$
centered at $\vc{a}_j$.
We assume that $B_r(\vc{a}_j)$  do not intersect. 

In our method, spirals are implicitly defined
by two functions, $u$ and $\theta$ as follows:
\begin{align}
 \label{lv form}
  \Gamma_t := \{ \vc{x} \in \overline{W} | \
  u(t,\vc{x}) - \theta (\vc{x}) = 2 \pi n,
 \ \mbox{for some integer } \ n \},
\end{align}
where $\overline{W}$ is union of the
sets $W$ and its boundary.
Correspondingly, we define the orientation of a spiral by
${\bf n} = - \frac{\nabla (u - \theta)}{|\nabla (u - \theta)|}$.
$\theta(\vc{x})$  is a pre-determined function of the form 
\begin{equation}
 \label{sheet structure}
 \theta (\vc{x})
  = \sum_{j=1}^N m_j \arg (\vc{x} - a_j).
\end{equation}
This function reflects the sheet structure
of the lattice of atoms with screw dislocations, 
and it was first proposed by Kobayashi \cite{Kobayashi:1990th}
to model spiral curves.
The constants $m_j$ define the strengths of the spiral centers: 
each strength is the difference between the stength, $m_j^+$, of counter-clockwise rotating spirals (that are attached to $\vc{a}_j$) 
and $m_j^-$ for clockwise rotating ones. 
\cite[Definition 3,5]{OTG:2015JSC} 

The function $u(t, \vc{x})$ is called an auxiliary function
to be approximated by solving a partial differential equation in $W$ 
with suitable initial and boundary conditions:
\begin{equation}
 \label{lv eq}
 u_t - v_\infty |\nabla (u - \theta)|
  \left\{ \rho_c 
  	\mathrm{div} \frac{\nabla (u - \theta)}{|\nabla (u - \theta)|} + 1
  \right\}
  = 0 
  \quad \mbox{in} \ (0,T) \times W,
\end{equation}
with an initial value condition
$u(0,\vc{x}) = u_0 (\vc{x})$ for $\vc{x} \in \overline{W}$
for a continuous function $u_0$ on $\overline{W}$
satisfying
\begin{equation}
 \label{lv init}
 \Gamma_0 = \{ \vc{x} \in \overline{W} | \ u_0 (\vc{x}) - \theta (\vc{x})
 = 2 \pi n \ \mbox{for an integer} \ n \}. 
\end{equation}
We impose the right angle condition
between $\Gamma_t$ and the boundary of $W$,
which is denoted by $\partial W$.
This condition is given as
\begin{equation}
 \label{lv nbc}
 \langle \vec{\nu}, \nabla (u - \theta) \rangle = 0
  \quad \mbox{in} \ (0,T) \times \partial W,
\end{equation}
where $\vec{\nu}$ is the outer unit normal vector field
of $\partial W$, and $\langle \cdot, \cdot \rangle$
denotes the usual inner product in $\mathbb{R}^2$.

A few remarks are in order.
First, the discontinuity of  $\theta$ does not cause any problem in \eqref{lv eq} since $\nabla\theta$ can be defined uniquely.
In fact, $\nabla \theta$ is well-defined on $\overline{W}$ as
\[
 \nabla \theta
 = \sum_{j=1}^N \frac{m_j}{|x - a_j|^2}
 (- x_2 + a_{j,2}, x_1 - a_{j,1})
\]
for $\vc{x}=(x_1, x_2)$ and $\vc{a}_j = (a_{j,1}, a_{j,2})$
by taking a branch of $\theta$ so that
it is smooth around $\vc{x}$.

Second, notice that $u_0$ satisfying \eqref{lv init}
is not unique even if $u_0$ is considered in
the space of continuous functions.
However, the uniqueness of $\Gamma_t$
for a given $\Gamma_0$ is established 
provided that $u_0$ is continuous and
the orientation of $\Gamma_0$ given by $u_0$
is the same \cite{Goto:2008hy}.
In order words,  $\Gamma_t$ depends only on $\Gamma_0$
and its orientation, and is independent of
the choice of the functions that embed it.
Initial data $u_0$ for the simulations
in this paper will be chosen 
as a constant or constructed from
a union of lines: see the preivous paper\cite{OTG:2015JSC}
for details of the construction.

\subsection{Growth rate of the surface}
\begin{figure}[htbp]
 \begin{center}
  \includegraphics[scale=1.0]{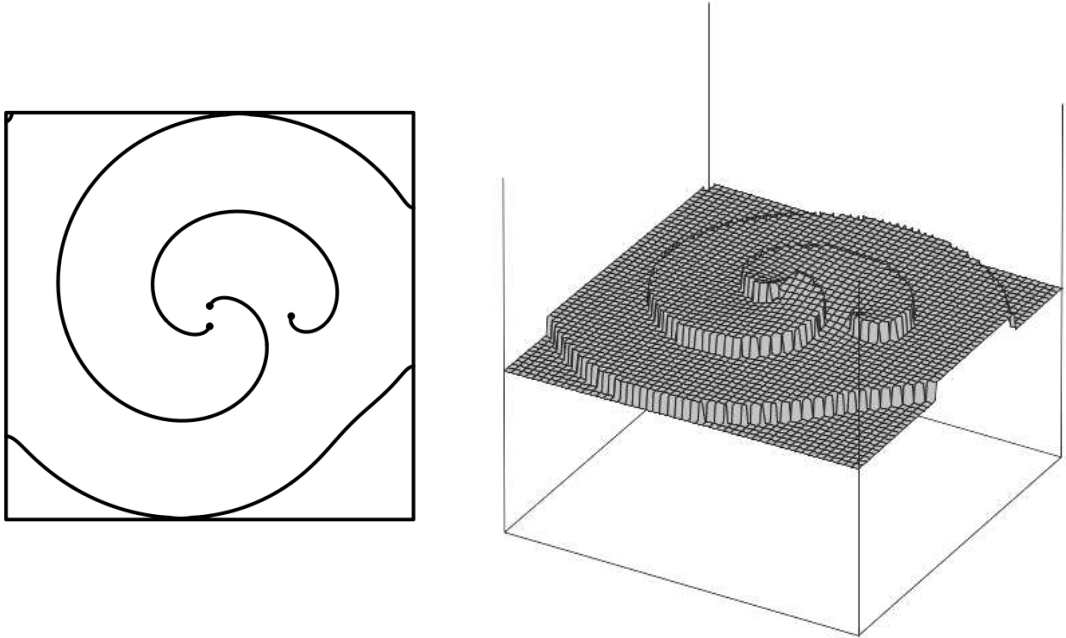}
  \caption{Example of level set for spirals
  and its height function.}
  \label{fig: height fct}
 \end{center}
\end{figure}

With given $\theta$ and $u$, $\Gamma_t$ is defined, and the height function  the growing crystal surface is defined as 
\begin{equation}
 h (\vc{x}) = \frac{h_0}{2 \pi} \theta_{\Gamma_t} (\vc{x}),
\label{eq:height}
\end{equation}
where $\theta_{\Gamma_t}$ is a branch of $\theta$
that has $2 \pi$-jump discontinuity
only on $\Gamma_t$. 
\cite{OTG:2015JSC}
See Figure \ref{fig: height fct}
for an example of $h(\vc{x})$
constructed from a level set for spirals. 
We define the mean growth height
in the time interval $[t_0,t]$ as
\[
 H(t;t_0) = \frac{1}{|W|}
 \int_W (h(t,\vc{x}) - h(t_0,\vc{x})) d\vc{x},
\]
where $|W|$ is the area of $W$.
Here and hereafter we shall use a notation
$H(t) := H(t;0)$
unless it is necessary to clarify
the initial time $t_0$.

The growth rate of the crystal surface is then given formally by
\begin{equation}
 R(t) = H'(t;t_0)
 = \frac{1}{|W|} \int_W h_t (t,\vc{x}) d\vc{x}.
\label{eq:growth-rate}
\end{equation}
However, $H(t)$
may not be differentiable somewhere and
may have oscillations with small amplitudes due to the domain shape.
Therefore, in this paper, we computed an ``effective''
growth rate of the crystal by a linear approximation 
that best fits, in the sense of least square,
the numerically computed values of $H(t_j)$ for $t_j$
in a chosen time interval. 
More precisely, we calculate $R_\triangle$ minimizing
\begin{equation}
 \label{line fitting}
  \min_{R_\triangle} \sum_{j=0}^K
  |H(t_0+j\Delta t,t_0) - R_\triangle
  (j \Delta t - t_0)|^2 
\end{equation}
with $\Delta t = (t_1 - t_0)/K$ for some $K \in \mathbb{N}$
on a time interval $[t_0, t_1]$.
Then, the coefficient
$R_\triangle$ gives the growth rate of the
crystal surface in $[t_0, t_1]$.

\section{New estimates of the growth rates and numerical results}

In this section, we discuss old and new estimates of crystal growth rates under different configurations of screw dislocations.
Our discussion is accompanied by the corresponding numerical simulations which serve both as motivation and verification
of the reported   new results.

\subsection{Discretization and numerical parameters}

We discretize \eqref{lv eq}--\eqref{lv nbc} on $W\subset\Omega=[-1,1]^2$  
with a finite difference scheme using the Cartesian grids
\[
 D_s = \{ (\frac{i}{100s}, \frac{j}{100s}) | \ -100s \le i,j \le 100s \}\subset\Omega=[-1,1]^2
\]
for  $s=1,2$, or $4$. Denote the grid spacing by $\Delta x=1/100s$.  
We solve the equation until $T=1$ using step size 
$\Delta t := \Delta x^2 / 10$.
The spiral centers $\vc{a}_1, \ldots, \vc{a}_N$ are chosen from $D_s$ and
$r < \Delta x$.
We calculate \eqref{lv eq}, \eqref{lv nbc} by the explicit finite
difference scheme of the form
\[
 u^{k+1}_{i,j} = u^k_{i,j}
 + v_\infty (\textup{I}^k_{i,j} + \rho_c \textup{II}^k_{i,j}),
\]
where $u^k_{i,j} = u(k \Delta t ,i \Delta x, j \Delta x)$ and
\begin{align*}
 \textup{I}^k_{i,j}
 & = \sqrt{|\tilde{\partial}_x (u - \theta)^k_{i,j}|^2 + |\tilde{\partial}_y (u - \theta)^k_{i,j}|^2}, \\
 \textup{II}^k_{i,j} & = \sqrt{|\hat{\partial}_x (u - \theta)^k_{i,j}|^2 + |\hat{\partial}_y (u - \theta)^k_{i,j}|^2}
 \left[ \mathrm{div} \frac{\nabla (u - \theta)}{|\bar{\nabla} (u - \theta)|} \right]^k_{i,j}.
\end{align*}
We refer the previous paper
by the authors \cite[\S 3.1]{OTG:2015JSC} for details
of the difference formulae $\tilde{\partial}_x w$, $\hat{\partial}_x w$,
and $\mathrm{div} (\nabla w/|\bar{\nabla} w|)$ for $w = u - \theta$.
Note that in the formula of $\tilde{\partial}_x w$
in the previous paper \cite[\S 3.1]{OTG:2015JSC},
the coefficient $\delta (=\Delta x)$ in front of $\mu$ is missing.

In this section we calculate the equation
\eqref{lv eq} with $v_\infty = 6$ and various different values of $\rho_c$
to obtain the evolution of spiral steps,
i.e., spiral steps evolves by
\[
 V = 6 (1 - \rho_c \kappa) 
\]
with some $\rho_c$ for verifying our speculations.
We also set $h_0 = 1$.

\subsection{Single spiral}

As the first test, we consider a situation where a single screw dislocation
providing a single spiral step
with the height of an atom.
We call such a step \textit{a unit spiral step},
and such a situation \textit{a single spiral} case.

Burton et al. \cite{Burton:1951tr} pointed out that
the growth rate of the crystal surface
by a steadily rotating unit spiral step is
\[
   R^{(0)} = \frac{\omega h_0}{2\pi},
\]
where $\omega$ is the angular velocity of the rotating spiral.
They estimated that $\omega = \omega_1 v_\infty / \rho_c$,
and $\omega_1 = 1/2$
with an approximation by an Archimedean spiral,
or $\omega_1 = \sqrt{3}/[2(1 + \sqrt{3})] \approx 0.315$
with an improved approximation.
Cabrera and Levine \cite{Cabrera-Levine:1956}
estimated that $\omega_1 = 2\pi / 19 \approx 0.330694$,
and this number was referred to in
Miura--Kobayashi's paper \cite{Miura:2015CGD}.
Ohara and Reid \cite{Ohara:1973ag}
proposed to solve an ordinary differential equation
in a half line to construct a spiral in $\mathbb{R}^2$.
They use the shooting method to construct a solution
and calculate $\omega_1$ numerically as a shooting
parameter. They obtained $\omega_1 = 0.330958061$.
\emph{In this paper, we assume that this quantity is more accurate physically
 and will use it as a reference in the following discussion.}
We compare our computation to the angular velcity obtained by Ohara and Reid:
\begin{equation}
 \label{GR formula: single}
  R^{(0)} = \frac{\omega_1 v_\infty h_0}{2 \pi \rho_c},
  \quad
  \omega_1 = 0.330958061.
\end{equation}

\begin{figure}[htbp]
 \begin{center}
  \includegraphics[scale=1.0]{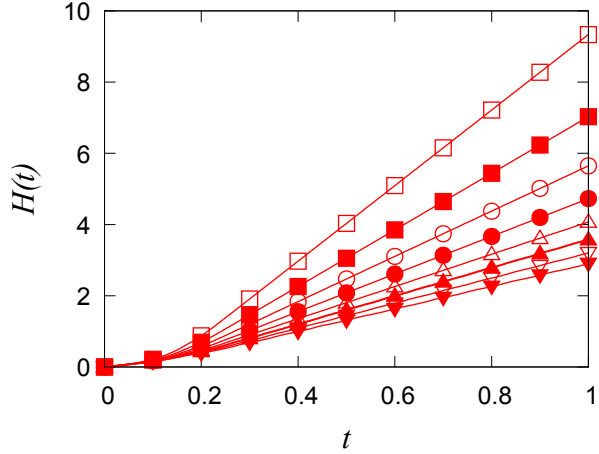}
  \caption{Graphs of $H (t)$
  for the evolution with
  a single screw dislocation and
  a unit spiral step.
  The horizontal axis means time $t$.
  Each line with a mark means the case
  $\rho_c = 0.03(\square)$,
  $0.04(\blacksquare)$,
  $0.05(\circ)$,
  $0.06(\bullet)$,
  $0.07(\triangle)$,
  $0.08(\blacktriangle)$,
  $0.09(\triangledown)$
  and
  $0.10(\blacktriangledown)$, respectively.} 
  \label{heights-single}
 \end{center}
\end{figure}

In the simulations, we set $N=1$, $m_1 = 1$, $\vc{a}_1 = 0$, and
\[
 \theta (x) = \arg x.
\]
Initial step is chosen as $\Gamma_0 = \{ (r,0) \in \overline{W}| \ r > 0 \}$.
In all of the evolutions presented in this paper, 
the height seems to grow linearly for $t \ge 0.3$.
Figure \ref{heights-single} presents the computed height $H(t) = H(t;0)$
with  $\rho_c $ ranging from $0.03$ to $0.1$.
We denote by $R_\triangle$ the growth rate
obtained from least square approximation of the computed height
in the time interval $[0.3,1.0]$.
Table \ref{table3-1}  shows some results
comparing $R_\triangle$ to $R^{(0)}$. 
We observe that the normalized differences
$e^{(0)} :=|R_\triangle - R^{(0)}|/R^{(0)}$
decrease at a rate which is larger than first order
in $\Delta x$.

Hereafter, we shall refer the above case
($N=1$, $m_1 = 1$, $\vc{a}_1 = 0$) or results
as \emph{a unit spiral} case.
\begin{table}[htbp]
\begin{center}
 \caption{Normalized differences $e^{(0)}$
 from numerical growth rates to the theoretical values
 by a unit screw dislocation.
 } \label{table3-1}\medskip
\begin{tabular}{|c||c|c|c|}
 \hline
 & \multicolumn{3}{c|}{$e^{(0)}$} \\
 \cline{2-4}
 $\rho_c$ 
 & $s=1$ & $s=2$ & $s=4$ \\ 
 \hline
 $ 0.030 $ 
	 & $0.006807$ & $0.004024$ & $0.001630$ \\ 
 \hline
 $ 0.040 $ 
	 & $0.005830$ & $0.002902$ & $0.001064$ \\ 
 \hline
 $ 0.050 $ 
	 & $0.005093$ & $0.002164$ & $0.000738$ \\ 
 \hline
 $ 0.060 $ 
	 & $0.004021$ & $0.001619$ & $0.000542$ \\ 
 \hline
 $ 0.070 $ 
	 & $0.003464$ & $0.001281$ & $0.000395$ \\ 
 \hline
 $ 0.080 $ 
	 & $0.002875$ & $0.001056$ & $0.000349$ \\ 
 \hline
 $ 0.090 $ 
	 & $0.002585$ & $0.001044$ & $0.000428$ \\ 
 \hline
 $ 0.100 $ 
	 & $0.002128$ & $0.000665$ & $0.000144$ \\ 
 \hline
\end{tabular}
\end{center} 
\end{table}

\subsection{Co-rotating pair}
\label{sec: co-rotating pair}

In the following, we study the dynamics of
co-rotating pair of spirals and derive a 
new formula \eqref{growth rate by a group on a line}
for the growth rate for $N$ co-rotating spirals.
Burton et al. \cite{Burton:1951tr}
pointed out that the growth rate
by a pair of co-rotating screw dislocations
at $\vc{a}_1$ and $\vc{a}_2$
depends on the distance $d := |\vc{a}_1 - \vc{a}_2|$
between the two screw dislocations.
(Here we have interpreted ``activity of screw dislocations''
in the classical paper by Burton et al.\cite{Burton:1951tr}
by ``growth rate'' on the above.
Hereafter, we similarly continue to use this interpretation.)
More precisely, 
\begin{enumerate}
 \item If the pair are far apart as
       $d > 2 \pi \rho_c =: d_c$,
       then the growth rate by the pair
       is indistinguishable from that of a unit spiral,
       i.e., $R^{(0)}$.
 \item If $d \ll \rho_c$,
       then the growth rate should be twice of
       $R^{(0)}$.
\end{enumerate}
On one hand they do not mention intermediate situations,
on the other hand
they estimated the growth rate of $N$ co-rotating
screw dislocations on a line with length $L$ as
\begin{equation}
 \label{activity by BCF}
  R^{(N)} (L) = \frac{N}{1 + L/(2 \pi \rho_c)} R^{(0)}.
\end{equation}
Our new formula gives a more accurate prediction of the critical distance separating the two cases mentioned above.

\begin{figure}[htbp]
 \begin{center}
  \includegraphics[scale=1.0]{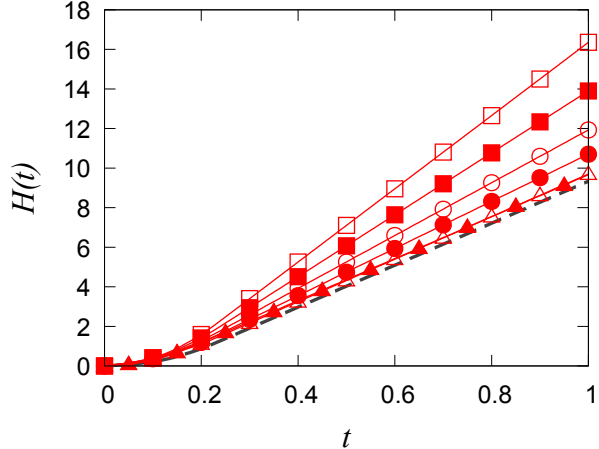}
  \caption{The left figure
  is the graphs of $H (t)$
  by a pair of co-rotating
  screw dislocations with 
  $\rho_c = 0.030$.
  The line with $\square$
  means the case of $d := |\vc{a}_1 - \vc{a}_2| = 0.04$.
  Similarly,
  the line with $\blacksquare$,
  $\circ$, $\bullet$, $\triangle$,
  and $\blacktriangle$ means the
  case of $d = 0.08$, $0.14$, $0.20$, $0.30$
  and $1.00$, respectively.
  Note that the graphs with $d=0.30$ and $d=1.00$
  are agree with each other.
  The dashed line is $H(t)$ of the unit spiral with
  the same $\rho_c$.
  }
  \label{height-rate coro:0.03}\medskip
 \end{center}
\end{figure}
We first present a set of numerical simulations showing that the formula (3.2) is not accurate even for $N=2$. 
Let
\[
 \theta (\vc{x}) = \arg (\vc{x} - \vc{a}_1) + \arg (\vc{x} - \vc{a}_2)
\]
for a given pair $\vc{a}_1 = (- \alpha, 0), \vc{a}_2 = (\alpha, 0) \in \Omega$
with $\alpha > 0$.
Set $u_0 \equiv 0$, so that the initial steps are on
the opposite line seguments of the line through $\vc{a}_1$ and $\vc{a}_2$:
\[
 \Gamma_0 =
 \{ \vc{a}_1 + r (\vc{a}_1 - \vc{a}_2) \in \overline{W}| \ r > 0 \}
 \cup \{ \vc{a}_2 + r (\vc{a}_2 - \vc{a}_1) \in \overline{W}| \
 r > 0 \}.
\]
Figure \ref{height-rate coro:0.03}
shows the graphs of $H$ computed with $\rho_c = 0.03$ 
in which we have $2 \pi \rho_c \approx 0.188496$.
From the Figure, we observe that the curves corresponding to 
$d = 0.30$ and $1.00$ are very close
to that computed from a single spiral.
\emph{Furthermore, they are quite far from the curve corresponding to $d=0.2$
(filled circles ($\bullet$) in the figure).}
Since $d=0.2$ is larger than $2 \pi \rho_c$, the numerical simulations suggest that
the critical distance $d_c$ is larger than $2 \pi \rho_c$.
In fact, the fitting lines for $d=0.20, 0.30, 1.00$
and the unit spiral for $\rho_c = 0.03$ are 
\begin{itemize}
 \item $d=0.20$:
       $H(t) \approx 11.926788t - 1.220501$,
 \item $d=0.30$:
       $H(t) \approx 10.761760t - 1.061625$,
 \item $d=1.00$:
       $H(t) \approx 10.611018t - 0.943661$,
 \item unit: $H(t) \approx 10.606435t-1.271220$.
\end{itemize}
Miura and Kabayashi\cite{Miura:2015CGD} reported that they
also found similar discrepancy using their phase field model.
It is further pointed out, without providing an explicit formula, that the growth rate by a co-rotating pair is indistinguishable
from that of the unit spiral if $d \ge 3 \pi \rho_c$.

To clarify the cause of such discrepancy, 
we present here a heuristic derivation of \eqref{activity by BCF} with $N=2$,
and with it we propose an improved formula for the growth rate, 
as well as the critical distance $d_c$.
Note that, in the following we denote an angular velocity of a rotating spiral
with \eqref{step velocity} by $\omega = \omega_1 v_\infty / \rho_c$,
where $\omega_1$ is as in \eqref{GR formula: single}.
\begin{figure}[htbp]
 \begin{center}
  \includegraphics[scale=1.0]{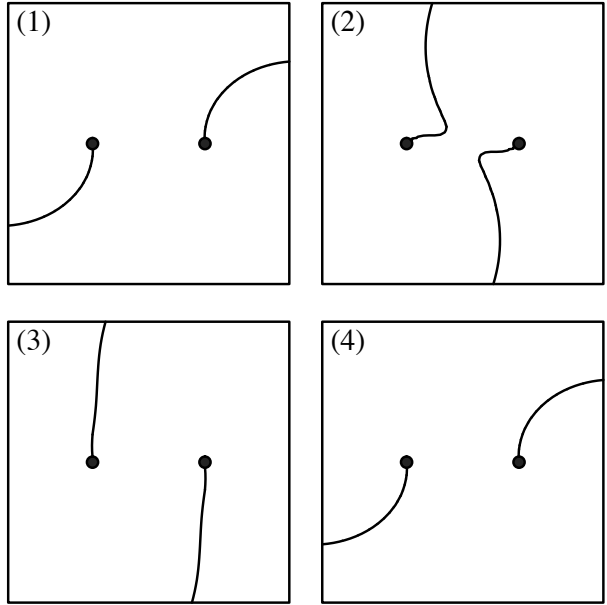}
  \caption{Process of rotation of co-rotating spirals.}
  \label{fig:coro_switch}
 \end{center}
\end{figure}
\begin{enumerate}
 \item The growth rate of
       a co-rotating pair with distance $d=|\vc{a}_1 - \vc{a}_2|$
       is given by
       \begin{equation}
	\label{formula: N-coro with T_L}
	 R^{(2)} (d) = \frac{2 h_0}{T_d},	
       \end{equation}
       where 
       $T_d$ is the time that
       the pair of spiral steps goes rotating around the pair.
 \item There are two fundamental motion 
       during the rotation of co-rotating spirals:
       switching spirals (from (1) to (3) in Figure \ref{fig:coro_switch})
       and half turn (from (3) to (4) in Figure \ref{fig:coro_switch}).
       Twice of the switchings and the half turns occur
       during the rotation once,
       and then 
       \[
	T_d = 2 (T_1 + T_2),
       \]
       where $T_1$ and $T_2$ is the time for
       the switching and the half turn.
 \item We regard the switching motion
       as the end point of spirals moves from $\vc{a}_1$ to $\vc{a}_2$
       with velocity $v_\infty$.
       Then, $T_1 = d / v_\infty$.
 \item In the half turn, 
       the angular velocity should be
       $\omega = \omega_1 v_\infty / \rho_c$.
       Then, $T_2 = \pi / \omega = \pi \rho_c / (\omega_1 v_\infty)$.
 \item Consequently we obtain
       \[
       T_d 
       = 2 \left( \frac{d}{v_\infty}
       +  \frac{\pi \rho_c}{\omega_1 v_\infty} \right)
       = \frac{2d + 2 \pi \rho_c / \omega_1}{v_\infty}.
       \]
       By combining 
       \eqref{formula: N-coro with T_L}, \eqref{GR formula: single}
       and the above we obtain
       \begin{equation}
	 R^{(2)} (d) 
	 = \frac{2}{1 + d \omega_1/(\pi \rho_c)}
	 \cdot \frac{v_\infty \omega_1 h_0}{2 \pi \rho_c}
	 = \frac{2}{1 + d \omega_1 / (\pi \rho_c)} R^{(0)}.
	 \nonumber
       \end{equation}
\end{enumerate}

Hence, for a pair of co-rotating spirals, 
we obtain the estimate of the growth rate
\begin{equation}
 \label{gr by a close co-rotating pair}
  R^{(2)} (d)
  = \frac{2}{1 + d \omega_1 / (\pi \rho_c)}
  R^{(0)}, \quad
  \omega_1 = 0.330958061,
\end{equation}
where $d$ is the distance between the two spiral centers which is assumed to be small.
Furthermore, since $R^{(2)} (d) < R^{(0)}$
if $d > \pi \rho_c / \omega_1$,
the growth rate with a co-rotating pair should be revised as
\begin{equation}
 \label{growth rate of a co-rotating pair}
  \tilde{R}^{(2)} (d)
  =
  \left\{
   \begin{array}{ll}
    R^{(2)} (d) & \mbox{if} \ d < \pi \rho_c / \omega_1, \\
    R^{(0)} & \mbox{otherwise}.
   \end{array}
  \right.
\end{equation}
Consequently, the critical distance is revised to 
\begin{equation}
 \label{revised critical distance}
  \tilde{d}_c = \frac{\pi \rho_c}{\omega_1}, \quad
  \omega_1 = 0.330958061.
\end{equation}
We remark that with $\omega_1=1/2$
the formulae \eqref{gr by a close co-rotating pair}
and \eqref{revised critical distance} reduce
to the predictions by Burton et al. \cite{Burton:1951tr}

\begin{figure}[htbp]
 \begin{center}
  \includegraphics[scale=1.0]{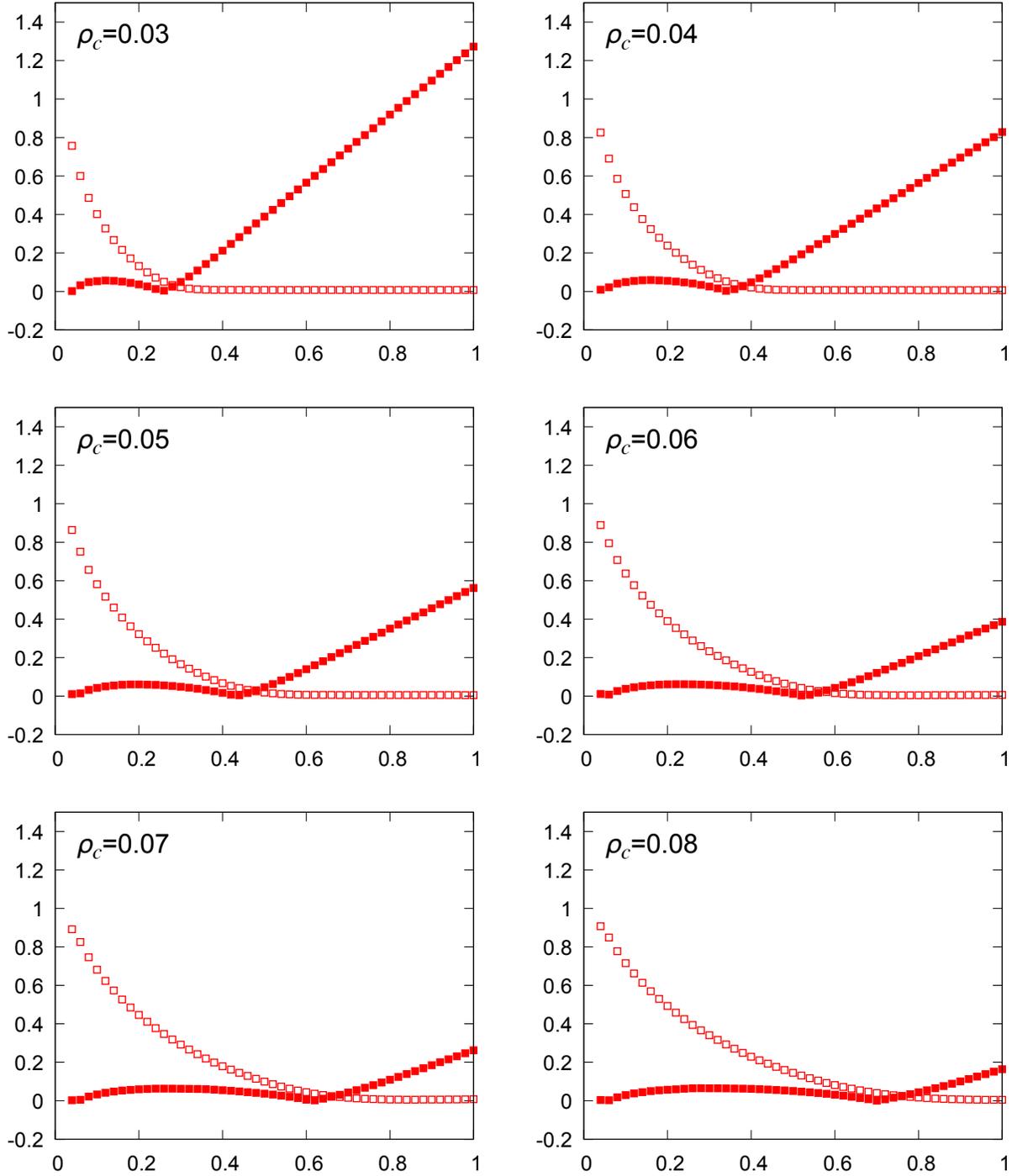} 
   \caption{Graphs of normalized differences
  $e^{(0)} (d) \ (\square)$
  and $e^{(2)} (d) \ (\blacksquare)$ for the pair
  $\vc{a}_1 = (-k \Delta x, 0)$, $\vc{a}_2 = (k \Delta x, 0)$
  with respect to the distance
  $d=|\vc{a}_1 - \vc{a}_2| = 2 k \Delta x$. }
  \label{errors co-sym}\medskip
 \end{center}
\end{figure}

For verification we report the normalized differences
\[
 e^{(0)} (d)
 := \frac{|R_\triangle (d) - R^{(0)}|}{R^{(0)}}, \quad
 e^{(2)} (d)
 := \frac{|R_\triangle (d) - R^{(2)} (d)|}{R^{(2)} (d)}
\]
with respect to the distance $d = |\vc{a}_1 - \vc{a}_2|$.
Again, $R_\triangle$ computed by solving in \eqref{line fitting}
with the numerical data on $t \in [0.3,1.0]$.
The numerical simulations are performed with
the centers
\[
 \vc{a}_1 = (- k \Delta x, 0), \quad \vc{a}_2 = (k \Delta x, 0)
 \quad (2 \le k \le 50),
\]
where $s=1$.
Figure \ref{errors co-sym} presents numerical results
of $e^{(0)} (d)$
and $e^{(2)} (d)$.
We observe that
$e^{(0)} (d)$ is small
if $e^{(2)} (d)$ is large,
and inversely $e^{(2)} (d)$ is small
if $e^{(0)} (d)$ is large.

From the numerical results we also can define
the numerical critical distance $\bar{d}_c$
dividing the co-rotating pair and independent
two single spirals as
\[
 \bar{d}_c = \sup \{d; \ e^{(2)} (d) < e^{(0)} (d) \}.
\]
From Figure \ref{errors co-sym} it seems that
$e^{(0)} (d)$ and $e^{(2)} (d)$
crosses only once in all the cases,
so that we now calculate $\bar{d}_c$
with linear interpolation;
\[
 \bar{d}_c \approx
 \frac{Y_1 d_{\bar{k}} + Y_0 d_{\bar{k}+1}}{Y_1 + Y_0},
\]
where $\bar{k}$
is such that
$e^{(2)} (d_{\bar{k}}) \le e^{(0)} (d_{\bar{k}})$
and 
$e^{(2)} (d_{\bar{k}+1}) > e^{(0)} (d_{\bar{k}+1})$
for $d_k = 2 k \Delta x$, and
\[
 Y_j = |e^{(0)} (d_{\bar{k} + j}) - e^{(2)} (d_{\bar{k} + j})|.
\]
The computed results are tabulated in Table \ref{table3-2}.
\begin{table}[htbp]
\begin{center}
 \caption{Comparison of the critical distances:
 $d_c=2 \pi \rho_c$ by Burton et al. \cite{Burton:1951tr},
 the revised distance
 $\tilde{d}_c = \pi \rho_c / \omega_1$, and the numerically observed 
 critical distance $\bar{d}_c$.}
 \label{table3-2}
\begin{tabular}{|c|r|r|r|}
 \hline
 \rule{0pt}{2.5ex}
$\rho_c$ & $2 \pi \rho_c$ & $\tilde{d}_c$ & $\bar{d}_c$ \\ 
\hline
$ 0.030 $ & $ 0.188496 $ & $ 0.284773 $ & $ 0.284813 $ \\ 
\hline
$ 0.040 $ & $ 0.251327 $ & $ 0.379697 $ & $ 0.379700 $ \\ 
\hline
$ 0.050 $ & $ 0.314159 $ & $ 0.474621 $ & $ 0.474650 $ \\ 
\hline
$ 0.060 $ & $ 0.376991 $ & $ 0.569545 $ & $ 0.569574 $ \\ 
\hline
$ 0.070 $ & $ 0.439823 $ & $ 0.664469 $ & $ 0.664486 $ \\ 
\hline
 $ 0.080 $ & $ 0.502655 $ & $ 0.759394 $ & $ 0.759396 $ \\
\hline
 \end{tabular}
\end{center}
\end{table}

Note that the estimate \eqref{gr by a close co-rotating pair}
is still rough in the sense that 
\[
 e_{dist} = \frac{|\bar{d}_c - \tilde{d}_c|}{\tilde{d}_c}
\]
increase as $\Delta x$ decreases; see Table \ref{table3-3}.
On the other hand, one finds that 
$e^{(0)}$, the normalized difference between
the computed rate and the reference rate of a single spiral, 
approaches 1 as  $d \to 0$.
The limiting case corresponds to $d = 0$ and
$\theta (\vc{x}) = 2 \arg \vc{x}$ is considered, 
and  it is proved that
the growth rate of the surface is $2 R^{(0)}$
if the two spirals agree with each other up to a rotation.
\cite{Ohtsuka:2014suriken}

The numerical growth rates obtained
in this subsection will be refered as $R^{(2)}_\triangle$
in the following sections.

\begin{table}[htbp]
 \begin{center}
  \caption{Normalized differences of the critical distance
  between $\tilde{d}_c$ and $\bar{d}_c$.}
  \label{table3-3}\medskip
  \begin{tabular}{|c|c|c|}
   \hline
   & \multicolumn{2}{|c|}{$e_{dist}$} \\
   \cline{2-3}
   $\rho_c$ & $s=1$ & $s=2$ \\
   \hline
   $ 0.030 $ & $ 0.000143 $ & $ 0.000275 $ \\
   \hline
   $ 0.040 $ & $ 0.000008 $ & $ 0.000013 $ \\
   \hline
   $ 0.050 $ & $ 0.000062 $ & $ 0.000108 $ \\
   \hline
   $ 0.060 $ & $ 0.000051 $ & $ 0.000094 $ \\
   \hline
  \end{tabular} 
 \end{center}
\end{table}

\subsection{Pair with opposite rotations}

Consider the case that
there is a pair of unit screw dislocations
with opposite rotation.
Burton et al. \cite{Burton:1951tr} pointed out
on this case as follows.
\begin{enumerate}
 \item If $d=|\vc{a}_1 - \vc{a}_2| < 2 \rho_c$ no growth occurs
       (called \textit{inactive pair}).
 \item If $d$ is around $3 \rho_c$,
       then the growth rate is about
       $1.1 \times R^{(0)}$.
 \item If $d \to \infty$,
       then the growth rate decays exponentially
       to $R^{(0)}$. 
\end{enumerate}
We shall verify the above speculations numerically;
in particular, on the estimate of the growth rate
with the above case (ii) 
and on the distance attaining the maximal growth rate.
In this section, we choose the initial step
as a line between $\vc{a}_1$ and $\vc{a}_2$:
\[
 \Gamma_0 = \{ \lambda \vc{a}_1 + (1 - \lambda) \vc{a}_2 \in \overline{W} | \ 0 \le \lambda \le 1 \}.
\]

\begin{figure}[htbp]
 \begin{center}
  \includegraphics[scale=1.0]{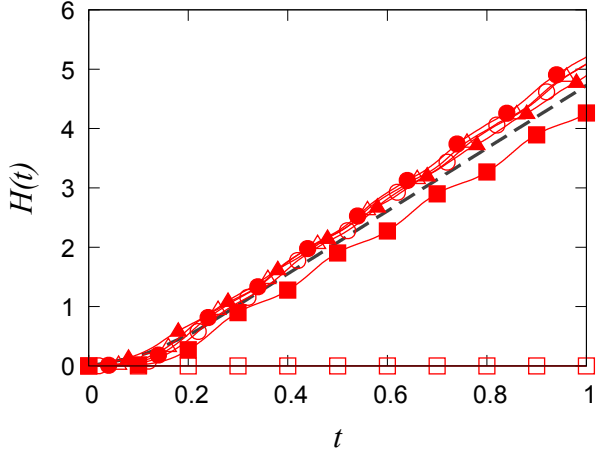}
  \caption{Graphs of $H (t)$ for
  a pair with opposite rotations with $\rho_c = 0.06$.
  Each graph means that $d=0.10<2 \rho_c$($\square$),
  $d=0.14 < 3 \rho_c$($\blacksquare$),
  $d=0.18 = 3 \rho_c$($\circ$),
  $d=0.24 = 4 \rho_c$($\bullet$),
  $d=0.36$($\triangle$) and $d=1.0$($\blacktriangle$).
  The dashed line denotes the graph of $H(t)$
  on the unit case spiral with $\rho_c = 0.06$.}
  \label{height-rate op:0.06}
 \end{center}
\end{figure}

We first show the typical examples of
the graphs of $H(t)$ for pairs with opposite rotations
in Figure \ref{height-rate op:0.06}.
We present the numerical results using 
$d=0.10 < 2 \rho_c$, with $\rho_c = 0.06$ on for case (i),
$d = 0.14 \in (2 \rho_c, 3 \rho_c)$ as the case between (i) and (ii),
$d = 0.18 = 3 \rho_c$ and
$d = 0.24 = 4 \rho_c$ as (ii),
and $d = 0.36, 1.0  \gg 3 \rho_c$.
The dashed line in Figure~\ref{height-rate op:0.06}
denotes the graph of $H(t)$ on the unit spiral with $\rho_c = 0.06$.
We find that the evolutions by the pair with opposite rotations
is faster than the unit spiral, except the cases when $d = 0.10$ and $0.14$.

\begin{figure}[htbp]
 \begin{center}
  \includegraphics[scale=1.0]{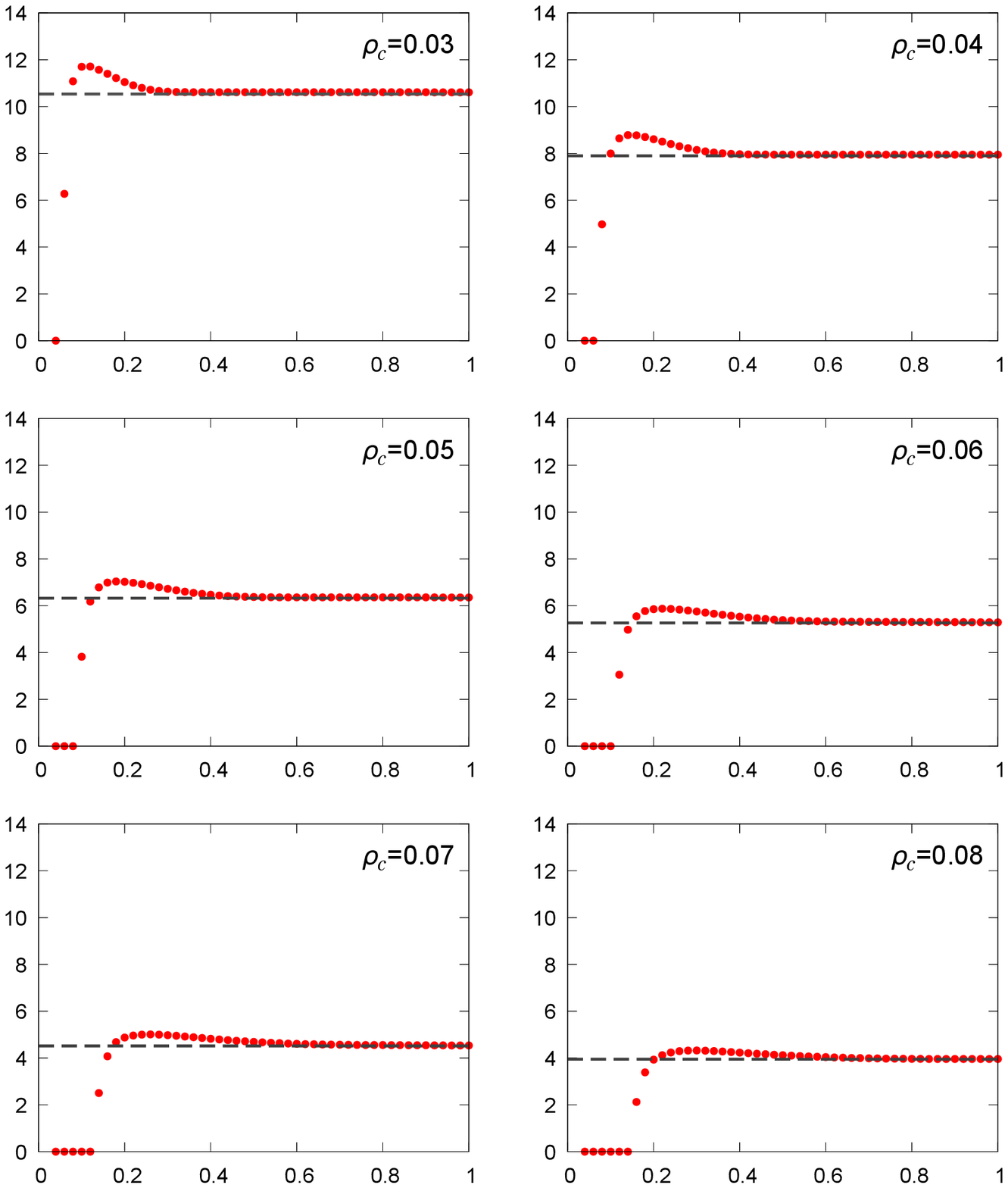}
  \caption{Graphs of $R_\triangle$ as a function of $d = |\vc{a}_1 - \vc{a}_2|$.
  The dashed line means $R^{(0)}$ for the each case.}
  \label{op_dist-rate}
 \end{center}
\end{figure}

To clarify the relation between $d=|\vc{a}_1 - \vc{a}_2|$
and $R_\triangle$,  
we numerically estimate the rates  $R_\triangle$ for several $\rho_c$ using 
computation performed in the time interval $[0.3,1.0]$.
In Fig~\ref{op_dist-rate} the red dots in each subplot correspond to 
$R_\triangle$ computed with centers
$\vc{a}_1 = (-k \Delta x, 0)$, $\vc{a}_2 = (k \Delta x, 0)$
for $ 2 \le k \le 50$ and  $\Delta x = 0.01$.
Each subplot shows the computations using a different $\rho_c$.
Thus, we observe that no growth occurs
when $d < 2 \rho_c$ for the each case.
By a similar argument
finding a stationary solution, \cite{Ohtsuka:OWR2010}
one can prove that 
no growth would occur at the critical distance
even if $d \le 2 \rho_c$.
However, due to numerical errors, we observed slow growth at this critical distance from our computations.
Our numerical simulations also show that, if $d$ is around $3 \rho_c$, 
the growth rate is
larger than that corresponding
to the unit spiral. In the subplots, the rate of the unit spiral is shown in the dashed lines.

\begin{figure}[htbp]
 \begin{center}
  \includegraphics[scale=1.0]{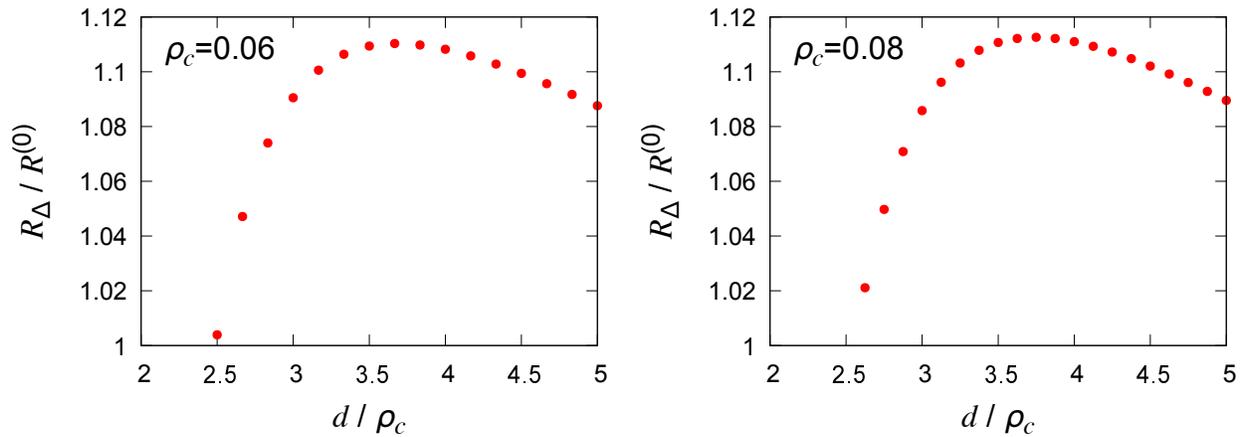}
  \caption{Graphs showing relation between $R_\Delta / R^{(0)}$
  and $d / \rho_c$ for $\rho_c = 0.06$ and $0.08$ with $s=2$.}
  \label{op_dist-rate_ratios}
 \end{center}
\end{figure}

In the column for $s=1$ in Table~\ref{op_max-dist}
we list the distance $d^*$
at which the growth rate attains its maximum,
and the normalized distance
$e^{(0)} = |R_\triangle - R^{(0)}|/R^{(0)}$
between $R_\triangle$ and $R^{(0)}$ at $d = d^*$.
We find $e^{(0)}$ is around $0.1$,
and the maximum growth rate
around $1.1 \times R^{(0)}$,
agreeing with the predictions 
by Burton et al.\cite{Burton:1951tr} or Miura--Kobayashi \cite{Miura:2015CGD}.
However, we also find that
the all results of $d^* / \rho_c$ examined here
are between $3.5$ and $4$,
which are larger than that value by 
Burton et al. \cite{Burton:1951tr}
or Miura--Kobayashi \cite{Miura:2015CGD}.
See also Figure \ref{op_dist-rate_ratios},
which shows the relation between $R_\triangle / R^{(0)}$
and $d / \rho_c$ for $\rho_c = 0.06$ and $0.08$ with $s=2$.
\begin{table}[htbp]
 \begin{center}
  \caption{{The distance between a pair of centers that result the maximal growth rate
  on a pair with opposite rotations.}}
  \label{op_max-dist}
  \medskip
 \begin{tabular}{|c|r|r|r|r|}
  \hline
  & \multicolumn{2}{|c|}{$s=1$} 
  & \multicolumn{2}{|c|}{$s=2$} \\ 
\hline
$\rho_c$ & $d^*$ & $e^{(0)}$ & $d^*$ & $e^{(0)}$\\ 
\hline
$ 0.030 $ & $0.120$ & $0.111637$ & $0.120$ & $0.083197$ \\ 
\hline
$ 0.040 $ & $0.140$ & $0.111348$ & $0.150$ & $0.107062$ \\ 
\hline
$ 0.050 $ & $0.180$ & $0.113245$ & $0.180$ & $0.108329$ \\ 
\hline
$ 0.060 $ & $0.220$ & $0.114888$ & $0.220$ & $0.110311$ \\ 
\hline
$ 0.070 $ & $0.260$ & $0.108985$ & $0.260$ & $0.104386$ \\ 
\hline
$ 0.080 $ & $0.300$ & $0.094041$ & $0.300$ & $0.112575$ \\ 
\hline
 \end{tabular} 
\end{center}
\end{table}

\subsection{Group on a line}

In this section we consider a situation where co-rotating
screw dislocations $\vc{a}_1, \ldots, \vc{a}_N$ with a unit spiral step
is ordered on a line, i.e., there exists $\lambda_j$ such that
$0 = \lambda_1 < \lambda_2 < \cdots < \lambda_N = 1$ and
$\vc{a}_j = (1 - \lambda_j) \vc{a}_1 + \lambda_j \vc{a}_N$.
Burton et al. \cite{Burton:1951tr} estimated the growth rate by
such a $\vc{a}_1, \ldots, \vc{a}_N$ as \eqref{activity by BCF}
if $|\vc{a}_{j+1} - \vc{a}_j| < d_c$ for each $j=1,\ldots, N-1$
and $|\vc{a}_1 - \vc{a}_N| = L$.
Then, 
by similar argument to obtain \eqref{gr by a close co-rotating pair}
as in 
the previous subsection ``co-rotating pair'',
we obtain the improved estimate of \eqref{activity by BCF}
as
\begin{equation}
 \label{growth rate by a group on a line}
  R^{(N)} (d) 
  = \frac{N}{1 + L \omega_1 / (\pi \rho_c)} R^{(0)},
  \quad \omega_1 = 0.330958061.
\end{equation}
We here remark that 
the estimate \eqref{growth rate by a group on a line}
by Burton et al. \cite{Burton:1951tr}
is independent of the distribution of $\vc{a}_j$'s on the line.
Miura--Kobayashi \cite{Miura:2015CGD} investigated the consistency of
the above formula and presented numerical simulations
for several co-rotating screw dislocations
($N \ge 2$) with $\omega_1 = 2 \pi/19$ and
equally arranged dislocations.
However, actually the distribution of screw dislocations
has influence to the growth rate.
We present below numerical results
verifying this assertion.

\begin{figure}[htbp]
 \begin{center}
  \includegraphics[scale=1.0]{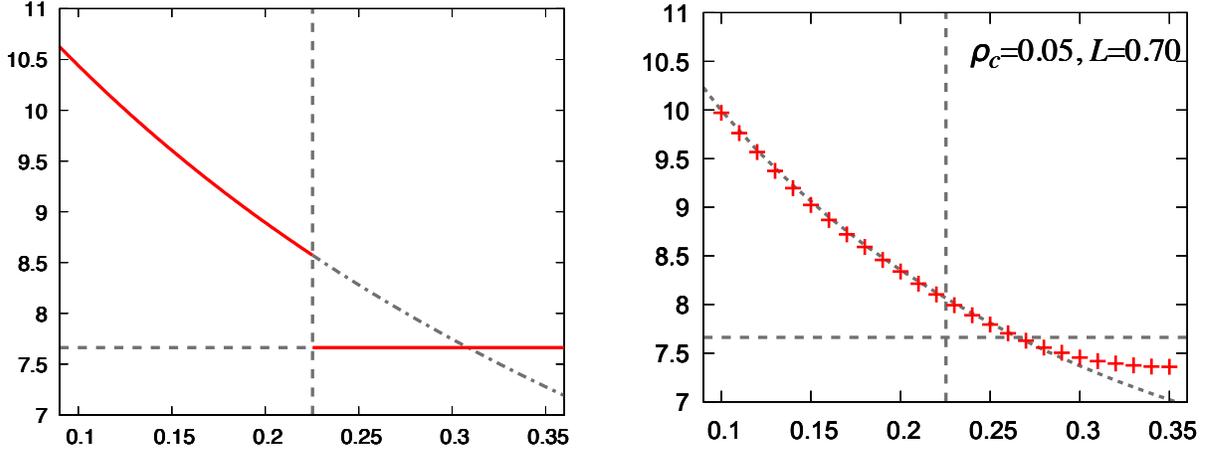}
  \caption{Comparison between
  estimates by Burton et al.\cite{Burton:1951tr} (left figure)
  and numerical simulations (right figure) with $\rho_c = 0.05$.
  The horizontal axis corresponds to  $|\vc{a}_1 - \vc{a}_2|$
  and the vertical axis is the rate.
  The dashed vertical line shows the distance $|\vc{a}_1 - \vc{a}_2| = 0.70 - \tilde{d}_c$
  (i.e., $|\vc{a}_2 - \vc{a}_3| = \tilde{d}_c$)
  with $\rho_c = 0.05$.
  The dashed horizontal line shows the rate $R^{(3)} (0.70) \approx 7.662046$ given 
  by \eqref{growth rate by a group on a line}.
  The chain line and the dotted line
  denote $R^{(2)}$
  (left figure) and $R^{(2)}_\triangle$
  obtained in
  the previous subsection ``co-rotating pair'',
  respectively.}
  \label{BCF triplets rate}
  \end{center}
\end{figure}

Consider the situation $N=3$ and
$\tilde{d}_c < |\vc{a}_1 - \vc{a}_3| < 2 \tilde{d}_c$,
for example,
\begin{equation}
 \label{location of triplets}
  \vc{a}_1 = (-0.35, 0), \quad \vc{a}_2 = (-k \Delta x, 0), \quad \vc{a}_3 = (0.35, 0) 
\end{equation}
with $\rho_c = 0.05$ for $k \ge 0$.
Note that the critical distance $\tilde{d}_c = 0.474650 $ is less than  distance
between the two farthest center $L=|\vc{a}_1 - \vc{a}_3| =0.70$.
Here we have used the revised critical distance as presented in 
\eqref{revised critical distance}.
In this case, the situations are divided into the following two situations.
\begin{itemize}
 \item[(a)] A group of triplets, if $|\vc{a}_2 - \vc{a}_3| \le \tilde{d}_c$,
 \item[(b)] A co-rotating pair and independent unit spiral, if $|\vc{a}_2 - \vc{a}_3| > \tilde{d}_c$.
\end{itemize}
Burton et al. \cite[\S 9.2]{Burton:1951tr} also pointed out that
the resultant growth rate is always that of the most active
independent group.
This suggests that the growth rate of  case (b) should be
$R^{(2)} (|\vc{a}_1 - \vc{a}_2|)$.
However, if the estimate by Burton et al. \cite{Burton:1951tr} were valid,
the growth rate by this group with respect to $|\vc{a}_1 - \vc{a}_2|$ would have a
unnatural discontinuity at $|\vc{a}_1 - \vc{a}_2| = L - \tilde{d}_c$
as in left figure of Figure \ref{BCF triplets rate}.
Hence, we examine the growth rate
of triplets at \eqref{location of triplets}
with $\rho_c = 0.05$, aiming at revealing whether or not 
such a discontinuity appears.
Our results are presented in the right plot in
Figure \ref{BCF triplets rate}.
Note that the initial data is chosen as $u_0 (\vc{x}) = 0$.

\begin{figure}[htbp]
 \begin{center}   
  \includegraphics[scale=1.0]{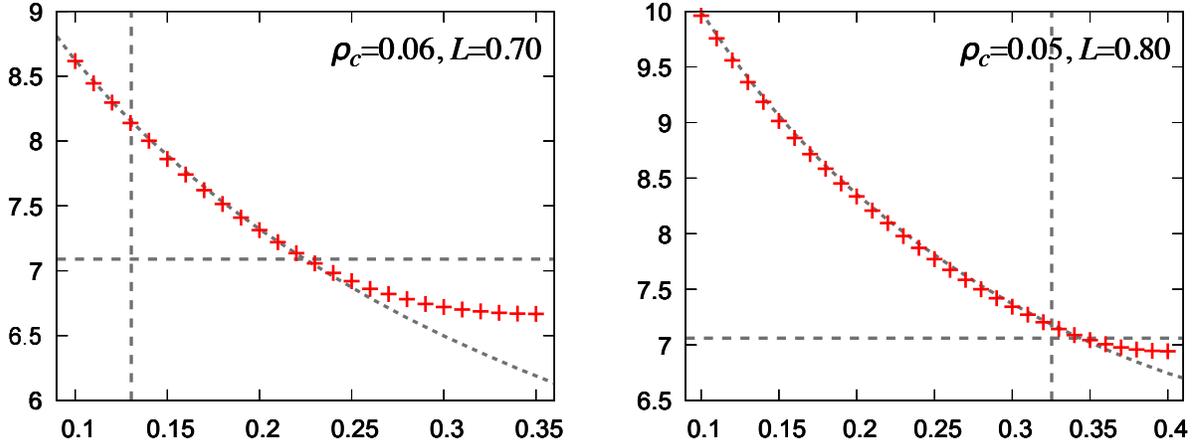}
  \caption{{Growth rates
  of the triplets with case (i) and (ii).
  The horizontal and vertical dashed line
  respectively denotes $R^{(3)} (L)$
  and $|\vc{a}_1 - \vc{a}_2| = L - \tilde{d}_c$
  for each cases.}}
  \label{dist-rate:3groups}
 \end{center}
\end{figure}

The right figure in Figure \ref{BCF triplets rate}
presents numerical results of the growth rate
by \eqref{location of triplets} with $\rho_c = 0.05$
for $0 \le k \le 25$ with respect to $d := |\vc{a}_1 - \vc{a}_2| = 0.35 - k \Delta x$.
The growth rates are estimated in the time interval $[0.3,1.0]$.
We also plot $R^{(2)}_\triangle (|a_1 - a_2|)$
as the dotted line. 
The growth rates of the triplets
follows closely the values of $R_\triangle^{(2)} (|\vc{a}_1 - \vc{a}_2|)$
on the region where $R^{(2)}_\triangle (|\vc{a}_1 - \vc{a}_2|) > R^{(3)} (L)$
even if the centers are sufficiently close
to be regarded as the group in the sense
of Burton et al. \cite{Burton:1951tr}
(to the right of the dashed vertical line).
However, when $R^{(2)}_\triangle$ becomes smaller than $R^{(3)} (0.70)$ 
(indicated by the dashed horizontal line),
the growth rate of the triplets becomes larger than $R^{(2)}_\triangle$.
Results similar to the above are obtained with the following setups:
\begin{enumerate}
 \item $\rho_c = 0.06$,
       $\vc{a}_1 = (-0.35, 0)$, $\vc{a}_2 = (- k \Delta x, 0)$, $\vc{a}_3 = (0.35,0)$
       for $-25 \le k \le 0$,
 \item $\rho_c = 0.05$,
       $\vc{a}_1 = (-0.40, 0)$, $\vc{a}_2 = (- k \Delta x, 0)$, $\vc{a}_3 = (0.40,0)$
       for $-30 \le k \le 0$.
\end{enumerate}
See Figure \ref{dist-rate:3groups}.

\begin{figure}[htbp]
 \begin{center}
  \includegraphics[scale=1.0]{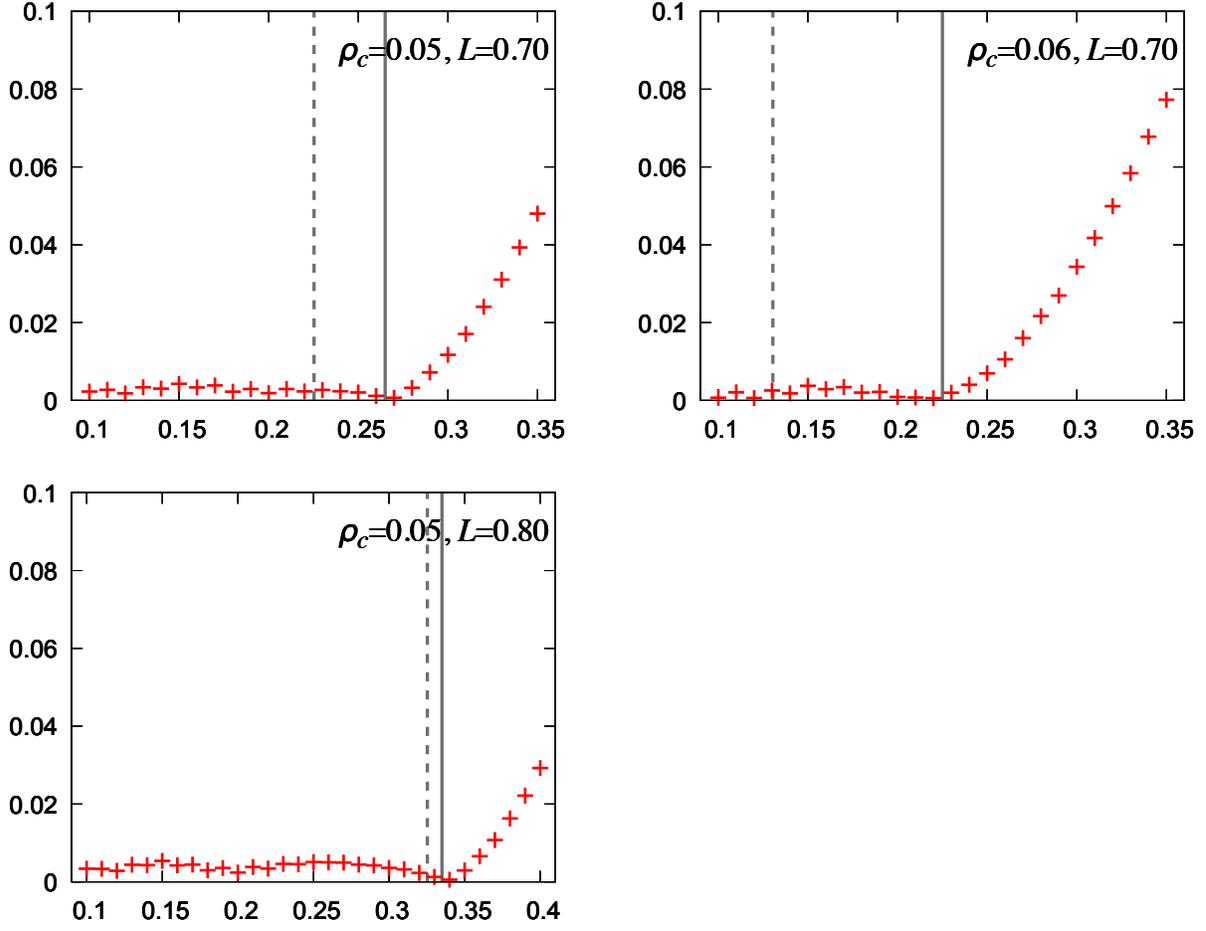}
  \caption{Normalized distances
  $|R_\triangle (d) - R^{(2)}_\triangle (d)|/R^{(2)} (d)$
  with $\rho_c = 0.05$, $L=0.7$,
  $\rho_c = 0.06$, $L=0.7$,
  and $\rho_c = 0.05$, $L=0.8$, respectively.
  The dashed lines are at $L-\tilde{d}_c$,
  and the solid lines at
  $|\vc{a}_1 - \vc{a}_2| = 0.265$ (top),
  $0.225$ (middle)
  or $0.335$ (bottom) approximately denote
  the distance where $R^{(2)}_\triangle$
  goes across $R^{(3)} (L)$ in each cases.}
  \label{dist-error:3groups}
  \end{center}
\end{figure}

We take the normalized distance
$|R_\triangle (d) - R^{(2)} (d)|/R^{(2)} (d)$ between
the numerical growth rates $R_\triangle (d)$ of the triplets
and $R^{(2)}_\triangle (d)$,
which is presented in Figure \ref{dist-error:3groups}.
Note that we choose $s=1$, i.e., $\Delta x = 0.01$
for the consistency of numerical results,
but we calculate $R_\triangle^{(2)} (d)$
with the linear interpolation
for $d = 0.11, 0.13, \ldots, 0.35$.
One can find that the growth rates are quite separated
from $R^{(2)}_\triangle$ if $d$ is larger than where
$R^{(2)}_\triangle$ goes across $R^{(3)} (L)$.
 
In the simulations with case (i),
note that the triplets 
should be regarded as a co-rotating group of
triplet if $k \le 21$ ($|\vc{a}_2 - \vc{a}_3| \le 0.56$).
However, the growth rate becomes quite larger
than that by a co-rotating pair
if $k \le 13$, where $R^{(2)} (d)$ is also smaller than
$R^{(3)} (0.7)$ with $\rho_c = 0.06$ if $d \ge 0.22$.
The case (ii) also proposes that
the growth rates become faster than those
by a co-rotating pair provided that $k \le 6$
although the triplets should be regarded
as a co-rotating group of triplet if $k \le 2$
($|\vc{a}_2 - \vc{a}_3| \le 0.38$).
One can find in the both cases that
the growth rate by a co-rotating triplet
is faster than a co-rotating pair,
however, slower than that calculated by
\eqref{growth rate by a group on a line}.

From our numerical simulations, we have the following
predictions {(note that the quantities are simulations by
\eqref{location of triplets} with $\rho_c = 0.05$):
\begin{itemize}
 \item The growth rate seems to decay smoothly
       for $d \ge 0.23$ although the triplets
       are classified as a group if $0.23 \le d \le 0.35$.
 \item The growth rate by the triplets
       becomes smaller than $R^{(3)} (0.70)$ if $d \ge 0.27$,
       however it continues to decay.
       Note that $R^{(2)}_\triangle (d)$ is also smaller than
       $R^{(3)} (0.70)$ if $d \ge 0.27$.
 \item The growth rate 
       is essentially larger than $R^{(2)}_\triangle (d)$
       if $d \ge 0.27$.
\end{itemize}
In summary, distribution of the screw dislocations on a line
influence to the growth rate of the whole group.
In particular, if the group can be regarded as 
sub groups of more closely positioned centers, then the resultant growth rate
shoud be that of the sub group with highest growth rate.
The quantity ${R^{(N)} (L)}$ possibly plays a role
of threshold changing the mode of the evolution.
However, we find no estimate for \eqref{location of triplets}
if ${ R^{(2)}_\triangle  (d)} < { R^{(3)} (L)}$.

As supplementary evidences to the above assertion, 
we present some examples of calculation 
of the growth rates for 4 co-rotating screw dislocations
as in the previous paper by the authors \cite{OTG:2015JSC}.
Recall the situation of the simulations:
4 co-rotating screw dislocations are located at
\[
 \vc{a}_1 = ( - a, 0), \ \vc{a}_2 = (-b,0), \
 \vc{a}_3 = (b,0), \ \vc{a}_4 = (a,0)
\]
with
\begin{enumerate}
 \item[(a)] $a=0.06$ and $b=0.02$, 
 \item[(b)] $a=0.15$ and $b=0.11$.
\end{enumerate}
The evolution equation is
\[
 V = 5 (1 - 0.02 \kappa),
\]
i.e., $v_\infty = 5$ and $\rho_c = 0.02$.
We choose the initial steps as
\begin{align*}
 \vc{a}_1: \quad & \{ \vc{a}_1 + (-r, 0) | \ r > 0 \}, \\
 \vc{a}_2: \quad & \{ \vc{a}_2 + (0, -r) | \ r > 0 \}, \\
 \vc{a}_3: \quad & \{ \vc{a}_3 + (0, r) | \ r > 0 \}, \\
 \vc{a}_4: \quad & \{ \vc{a}_4 + (r, 0) | \ r > 0 \}.
\end{align*}
The details of the initial data for these simulations,
and the profiles of spiral steps at $t = 0.5$
are given in the previous paper \cite{OTG:2015JSC}.
Profiles are slightly different from each other.
We now give a classification
if these situations are
a group of 4 co-rotating screw dislocations,
or 2 pairs from a view point of the growth rates.
\begin{figure}[htbp]
 \begin{center}
  \includegraphics[scale=1.0]{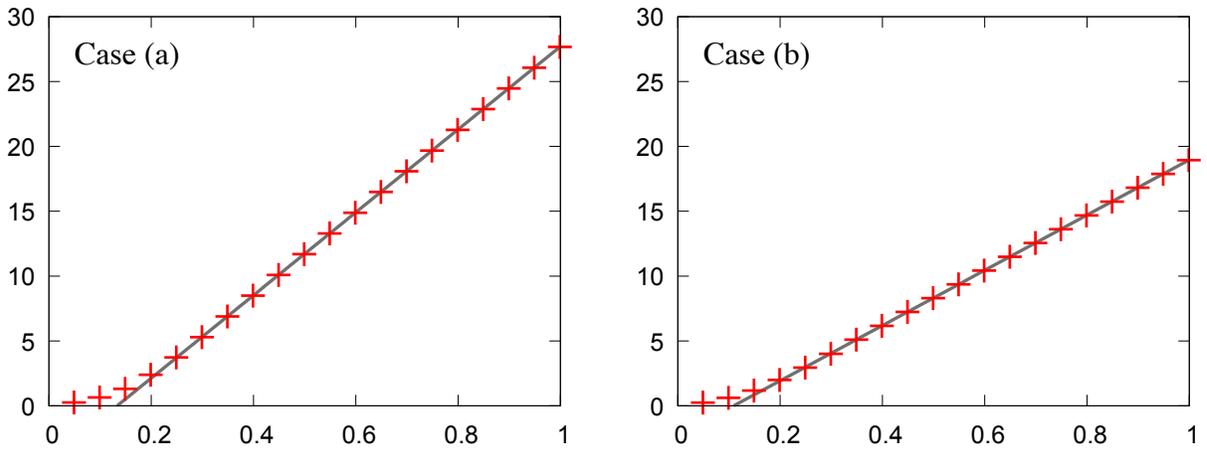}
  \caption{Graphs of $H(t)$ by
  4 screw dislocations (a)$(\pm 0.06,0)$
  and $(\pm 0.02,0)$(left),
  or (b)$(\pm 0.15, 0)$ and $(\pm 0,11, 0)$(right).
  The points denote the data of $H(t)$
  per the time span $\Delta t = 0.05$, and
  solid lines denote the fitting line by the data
  of $H(t)$ in $[0.3,1]$.}
  \label{height:4centers}
 \end{center}
\end{figure}
See Figure \ref{height:4centers} for the data plots
of $H(t)$ on these simulations.
Each fitting line is as follows;
\begin{enumerate}
 \item[(a)] $H(t) \approx 31.96154528 t - 4.26148724$,
 \item[(b)] $H(t) \approx 21.29516137 t - 2.33023461$.
\end{enumerate}
Then, the growth rate of the case (a) is
$R_{\triangle, 1} = 31.96154528$,
and that of the case (b)
is $R_{\triangle, 2} = 21.29516137$.
The growth rate $R^{(0)}$ with $v_\infty=5$, $\rho_c = 0.02$
is
\[
  R^{(0)} = \frac{5 \omega_1}{0.04 \pi} \approx 13.168403.
\]
The possibility of the classification (a) is
\begin{enumerate}
 \item[(a1)] a group of $\{ \vc{a}_1, \vc{a}_2, \vc{a}_3, \vc{a}_4 \}$ with length $L=0.12$,
 \item[(a2)] a group of $\{ \vc{a}_1, \vc{a}_2, \vc{a}_3 \}$ and an independent $\{ \vc{a}_4 \}$,
 \item[(a3)] a pair of pairs $\{ \vc{a}_1, \vc{a}_2 \}$ and $\{ \vc{a}_3, \vc{a}_4 \}$.
\end{enumerate}
Then, each growth rate is calculated as follows.
\begin{eqnarray*}
 & \mbox{(a1)} &
  {R^{(4)}} (0.12)
  = \frac{4}{1 + 0.12 \omega_1 / (0.02 \pi)} R^{(0)}
  \approx 32.273849,
  \\
 & \mbox{(a2)} &
  {R^{(3)}} (0.08)
  = \frac{3}{1 + 0.08 \omega_1 / (0.02 \pi)} R^{(0)}
  \approx 27.793385,
  \\
 & \mbox{(a3)} &
 R := \frac{2}{1 + 0.08 \omega_1 / (0.02 \pi)} R^{(2)} (0.04)
 \approx 30.608752.
\end{eqnarray*}

For case (a) one can find $R^{(4)} (0.12)$
is the closest to $R_{\triangle, 1}$.
So case (a) should be regarded as a group of four
co-rotating screw dislocations.
Case (b), on the other hand, 
should be regarded as two independent (non-interacting) pairs
$\{ \vc{a}_1, \vc{a}_2 \}$ and
$\{ \vc{a}_3, \vc{a}_4 \}$ since $\vc{a}_2$ and $\vc{a}_3$ are disconnected
in the sense that  $| \vc{a}_2 - \vc{a}_3 | = 0.22 > \tilde{d}_c$.
Thus, the growth rate should be estimated as $R^{(2)} (0.04) \approx 21.753470$.
Even if we regard case (b) as a group of four screw
dislocations on a line of length 0.30,
\eqref{growth rate by a group on a line} gives $R^{(4)} (0.30) \approx 20.414480$, which is
farther than $R^{(2)} (0.04)$.
Similarly, if we treat $\{ \vc{a}_1, \vc{a}_2 \}$
and $\{ \vc{a}_3, \vc{a}_4 \}$ as two effective pairs of centers,
we obtain 
\[
 R = \frac{2}{1+0.26\omega_1 / (0.02 \pi)}
 R^{(2)} (0.04) \approx 18.361125
\]
by calculation similar to (a3).
Note that the centers of the pair in this case
should be regarded as $(\pm 0.13, 0)$.

\subsection{Grouping of centers and the effective growth rate}

According to the classical paper by Burton et al. \cite{Burton:1951tr}  the
growth rate of crystal surface by several screw dislocations
could be estimated systematically by analyzing the rates of subgroups
of screw dislocations independently.
The procedure is summarized below:
\begin{itemize}
 \item Inactive pairs are disregarded.
 \item Dislocations are collected into ``disjoint'' subsets.
       We shall refer to each of such subsets as a group. 
       In each group, any dislocation center is no farther than
       $\tilde{d}_c$ 
       away from another dislocation center in the same group.
       On the other hand, subsets
       $\mathcal{A}$ and $\mathcal{B}$ of centers are
       disjoint if $|\vc{a} - \vc{b}| > \tilde{d}_c$
       for any $\vc{a} \in \mathcal{A}$ and $\vc{b} \in \mathcal{B}$.
       As in the previous sections, $\tilde{d}_c$  is the critical distance
       of a co-rotating pair.
 \item An effective center and strength is assigned to each group.
       Let $\vc{a}_1, \vc{a}_2, \ldots, \vc{a}_N$ be in a group.
       The strength $n$ of the group is defined as
       \begin{align*}
	n = m_1 + m_2 + \cdots + m_N,
       \end{align*}
       where
       $m_j=m_j^+ - m_j^-$, and $m_j^+$ (resp. $m_j^-$) is 
       the number of single spiral steps
       with counter-clockwise (resp. clockwise) rotational orientation
       associated with $\vc{a}_j$.
 \item 
       Each group's growth rate is
       approximated by the rate of the effective spiral center
       and its strength, which is roughly $\max(|n|,1)  R^{(0)}$.
       In particular, when $n=0$  the growth rate of the group is
       approximately the same as (and generally slightly greater than)
       $R^{(0)}$.
 \item The effective growth rate of the surface is then estimated by the maximum growth rate of the present groups. 
\end{itemize}

In this section, we study the validity of this procedure by numerical simulations involving 
a simplest setup that consists of three dislocation centers with opposite rotational orientations.
Through the numerical studies, we would like to carefully examine the aspects:
\begin{enumerate}
 \item Grouping of centers:
       whether the effective distance for grouping
       the centers is $\tilde{d}_c$ even for 
       pair of screw dislocations with opposite rotational orientations?
 \item Cancellation of the growth rate:
       which distance the cancellation of the growth rate
       by centers with opposite rotation
       occurs from?
       As we already see in the previous section that
       the growth rate by a pair with opposite rotation
       is approximately $1.1 \times R^{(0)}$
       if the distance of the pair is around $4 \rho_c < \tilde{d}_c$.
       But at such distances, Burton et al. 
       point out that the strength of the spiral pair is cancelled.
 \item Dependency of the growth rate on the distribution of centers.
       We think that non-smooth dependence on the center as a 
       consequence of the procedure is unnatural.
\end{enumerate}
%
%
%
For study of the cancellation issue,
we consider the evolution of the surface containing the centers
\begin{equation}
 \label{location of complex group}
 \vc{a}_1 = (0, -0.05; 1), \
 \vc{a}_2 = (0, 0.05; 1), \
 \vc{a}_3 = (k \Delta x, 0; -1), \
 \mbox{for} \ k \ge 0
\end{equation}
and the normal velocity of the steps prescribed by 
\begin{equation}
 \label{eq: complex group}
 V = 6 (1 - 0.04 \kappa),
\end{equation}
i.e., $v_\infty = 6$, $\rho_c = 0.04$.
Initial step is given as the following three lines:
\begin{align*}
 \vc{a}_1 : \quad & \{ \vc{a}_1 + (0,-r) | \ r > 0 \}, \\
 \vc{a}_2 : \quad & \{ \vc{a}_2 + (0,r) | \ r > 0 \}, \\
 \vc{a}_3 : \quad & \{ \vc{a}_3 + (r,0) | \ r > 0 \}.
\end{align*}
In \eqref{location of complex group}, we describe the screw dislocation $\vc{a}_j$ as the triplet $(p_j, q_j; m_j)$
where  $(p_j, q_j)$ is the coordinates of dislocation, and  $m_j \in \{ \pm 1 \}$ is the rotational orientation. 
According to the procedure described above, the growth rate can be estimated separately by the following three cases:
\begin{enumerate}
 \item[(c1)] If $|\vc{a}_1 - \vc{a}_3| \ge \tilde{d}_c$, 
	     the growth rate should be
	     $\max \{ R^{(2)} (0.10), R^{(0)} \}
	     = R^{(2)} (0.10)$.
 \item[(c2)] If $2 \rho_c \le |\vc{a}_1 - \vc{a}_3| < \tilde{d}_c$, 
	     then $\{ \vc{a}_1, \vc{a}_2, \vc{a}_3 \}$ are all in the same group
	     and $n=1$ on this group.
	     The growth rate should be
	     $1 \times R^{(0)} = R^{(0)}$.
 \item[(c3)] If $|\vc{a}_1 - \vc{a}_3| < 2 \rho_c := 0.08$, 
	     then $\vc{a}_1$ and $\vc{a}_3$ form an inactive pair,
	     and only $\vc{a}_2$ influences the growth rate. 
	     The growth rate should be $R^{(0)}$.
\end{enumerate}
Note that $|\vc{a}_1 - \vc{a}_3| = \sqrt{0.05^2 + (k \Delta x)^2}$,
so that (c1) applies when
\[
 k \Delta x \ge a^{**}
 := \sqrt{\tilde{d}_c^2 - 0.05^2} \approx 0.376390,
\]
and (c3) applies when
\[
 k \Delta x = a^*
 := \sqrt{(2 \rho_c)^2 - 0.05^2} \approx 0.062450.
\]
Hence, we obtain the estimate of the growth rate by
\eqref{location of complex group}
as in the top-left sub-figure of
Figure~\ref{BCF comptri estimate}.

\begin{figure}[htbp]
 \begin{center}
  \includegraphics[scale=1.0]{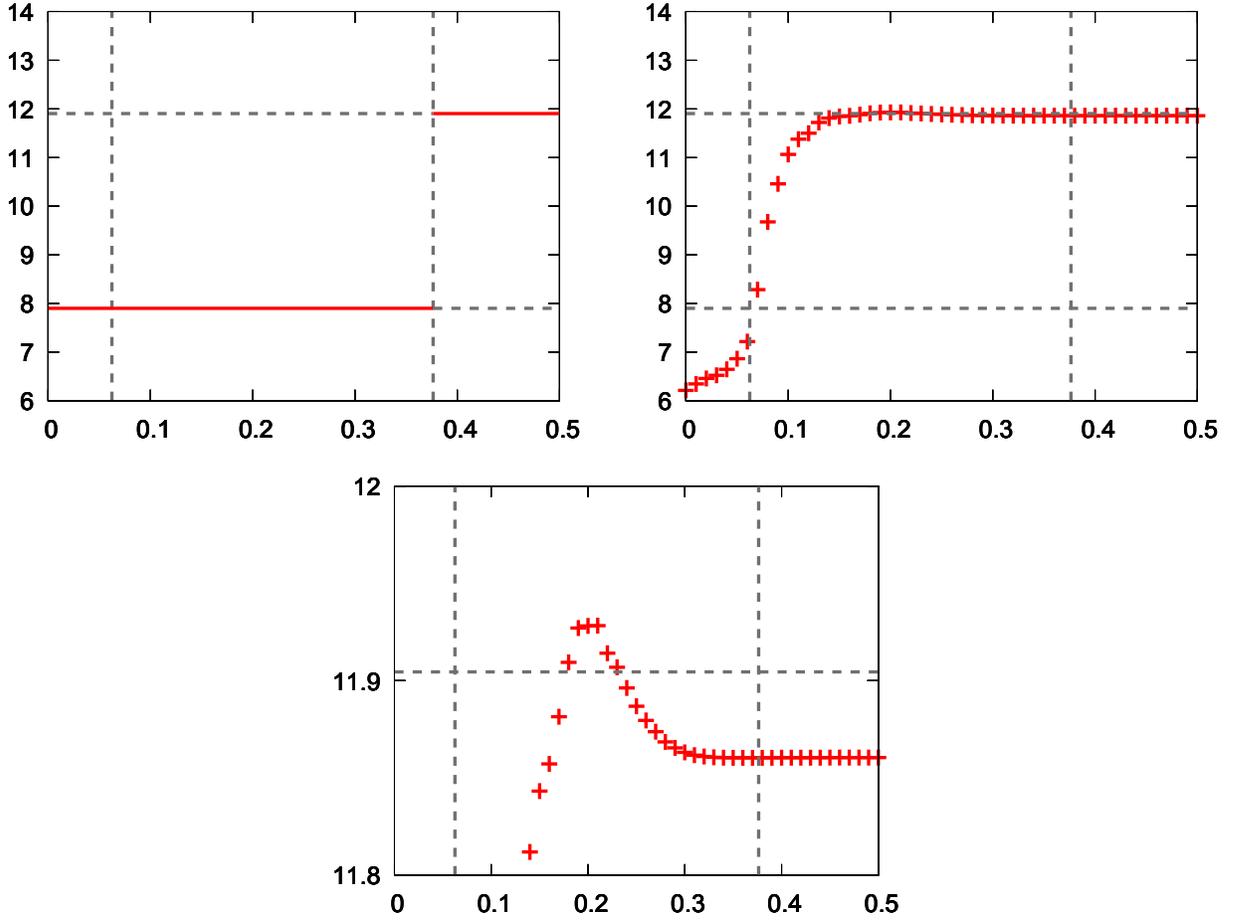}  
  \caption{The estimate by Burton et al. \cite{Burton:1951tr}
  of the growth rate by \eqref{location of complex group}
  (top-left)
  and its numerical results(top-right).
  The horizontal axis means $k \Delta x$,
  and the vertical dashed line are located at
  $k \Delta x = a^*$
  or $a^{**}$.
  In the bottom figure, we zoom in the numerical results  around 11.9 of $y$-axis.}
  \label{BCF comptri estimate}
 \end{center}
\end{figure}
The top-right sub-figure in
Figure \ref{BCF comptri estimate} shows the graph
of the numerical growth rate $R_\triangle$
in this situation,
which is calculated on a time interval $[0.7,2.0]$.
The horizontal axis means $k \Delta x$.
The horizontal dashed lines 
in the right figure are drawn at
$R^{(0)} \approx 7.901042$
and $R^{(2)}_\triangle (0.10) \approx 11.904457$.
The vertical dashed lines are drawn at
$k \Delta x = a^*$ and $k \Delta x = a^{**}$.
The bottom figure focusses the 
numerical results around 11.9 of $y$-axis for closer inspection.

The numerical results in Figure \ref{BCF comptri estimate} is summarized
as follows.
\begin{itemize}
 \item The growth rate keeps its quantity around
       $R^{(2)}_\triangle (|\vc{a}_1 - \vc{a}_2|)$
       until $k \Delta x \ge 0.15$.
       Note that $|\vc{a}_1 - \vc{a}_3| \approx 0.158114$
       ($k \Delta x = 0.15$)
       is close to $4 \rho_c = 0.16$.
 \item The growth rate is smaller than
       $R^{(0)}$ if $|\vc{a}_1 - \vc{a}_3| < 2 \rho_c$.
 \item The growth rate attains its maximum in $2 \rho_c < |\vc{a}_1 - \vc{a}_3| < \tilde{d}_c$,
       and monotonically decreases for larger values of $|\vc{a}_1 - \vc{a}_3|$.
\end{itemize}
The profile of the growth rate at $|\vc{a}_1 - \vc{a}_3| > 2 \rho_c$ looks like
the subplots in Figure \ref{op_dist-rate}.
So, similar overshooting of growth rates
as a pair with opposite rotational orientation
may appear if a group includes an accelerating pair with opposite rotations.
On the other hand, results reported in Figure 13, in particular the second bullet point above, 
implies that
an inactive pair in a group of screw dislocations may reduce
the growth rate of that group.

\begin{figure}[htbp]
 \begin{center}
  \includegraphics[scale=1.0]{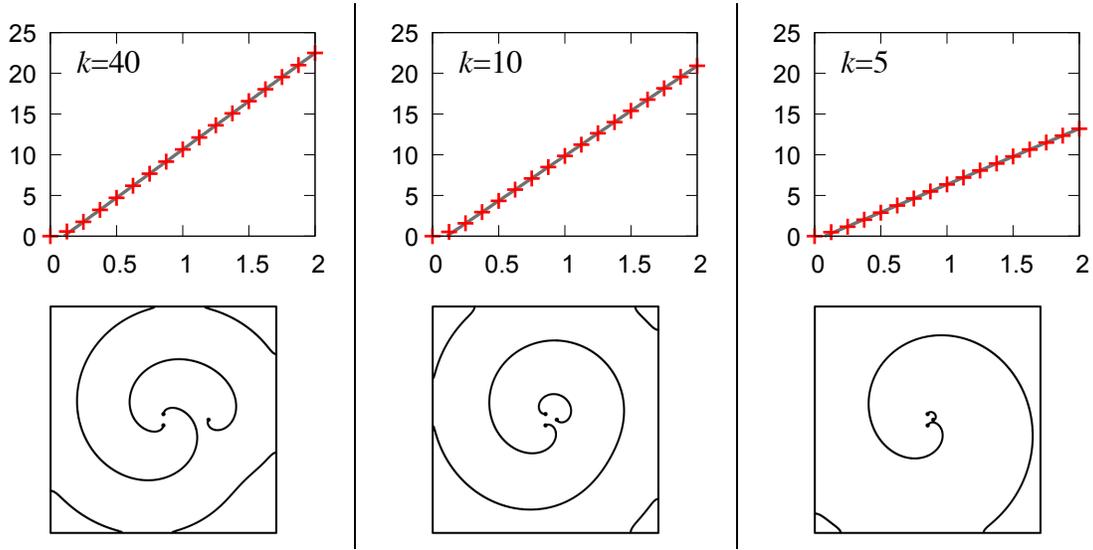} 
  \caption{Graphs of $H(t)$ and profiles of spirals at
  $t=2$ with $s=1$ ($\Delta x = 0.01$)
  for each case of (c1), (c2), (c3) as $k=40,10,5$, respectively.}
  \label{height:group2-1c}
 \end{center}
\end{figure}
In Figure \ref{height:group2-1c} we present three numerical simulations,
using $k=40, 10, 5$ corresponding to cases (c1)--(c3).
The numerically observed growth rates are, respectively,
\[
 R_\triangle = 11.860357, \quad
 R_\triangle = 11.064736, \quad
 R_\triangle = 6.863008.
\]
On the other hand, we have
\[
 R^{(0)} \approx 7.901042, \quad
 R^{(2)} (0.10) \approx 12.507902, \quad
 R^{(2)}_\triangle (0.10) \approx 11.904457,
\]
where $R^{(2)} (0.10)$ is calculated with
\eqref{gr by a close co-rotating pair},
and $R^{(2)}_\triangle (0.10)$ is the numerical result
obtained in 
the previous subsection ``co-rotating pair''.
There seem to be quite some discrepancy
between the presented computation and the ones 
predicted by Burton et al.\cite{Burton:1951tr}

Summarizing, 
in the numerical simulations presented above, we fix a co-rotating pair of spirals, and 
study the growth rate as the center of the third spiral, with opposite rotational orientation, approaches
the former two. 
We observed that if the distance, $L$, between the center of the third spiral and those of the co-rotating ones
is larger than the critical distance (for an inactive pair of spirals), then the growth rate
tends rapidly to the rate of the co-rotating pair, as $L$ becomes larger.
If $L$ is too small, then the growth rate is less than
that of the unit spiral.
In particular, a pair of
dislocation centers with opposite rotations in a group
should be regarded as a single center having approximately
the same growth rate as that of the unit spiral.

Regarding the issue of how centers should be grouped,
we propose the following steps in determining the effective growth rate:
\begin{enumerate}
 \item Each pair of centers
       $\vc{a}_i$ and $\vc{a}_j$
       with strength $m_i$, $m_j$ 
       should be regarded as a single center
       with the strength $m_i + m_j$ if $|\vc{a}_i - \vc{a}_j| < 2 \rho_c$.
       If $m_i + m_j = 0$, then the pair
       should be discarded from the surface.
 \item After the reduction described in (i), 
       groups of co-rotating pairs are identified by connecting those pairs centers of co-rotating 
       spirals that are within $\tilde{d}_c = \pi \rho_c / \omega_1$  distance to each other.
       The critical distance $\tilde{d}_c$ is defined in  \eqref{revised critical distance}.
 \item \label{num of spirals}
       If there is a group of co-rotating centers, denoted here as  $\mathcal{A}$,
       with the strength $n \ge 1$ and
       a center $\vc{b}$ with the negative strength $-m$
       with $m \ge 1$
       and they are close (for example, the distance
       between $\mathcal{A}$ and $b$ is less than $\tilde{d}_c$),
       then, they seem to be a single center with $|n - m|$
       spiral steps.
       Let $\mathcal{A}=\{ \vc{a}_j | \ j\in\mathcal{I}\}$ be a group of co-rotating spirals identified in (ii), 
       and $\vc{b}$ be a center outside of this group with negative strength
       $-m$ with $m \ge 1$. 
       If $|\vc{b}-\vc{a}_j|<\tilde{d}_c$ for some
       $\vc{a}_j \in \mathcal{A}$, then $\vc{b}$ and this group of co-rotating centers      
       should be considered as a group with which $|n-m|$ spirals are associated.
 \item \label{gr by group}
       If the center $\vc{b}$ is in the convex
       hull of $\mathcal{A}$ in (iii) and $n > m$,
       then the effective growth rate by $\mathcal{A}$ and $\vc{b}$
       should be reduced from that of $\mathcal{A}$ in a more elaborate fashion.
       Consider subgroups of $\mathcal{A}$.
       We call a subgroup \emph{$\vc{b}$-pure} if  
       $\vc{b}$ is outside of  the convex hull of this subgroup.
       The effective growth rate should be the maximum of the rates of $\vc{b}$-pure subgroups of
       $\mathcal{A}$ and $(n-m) \times R^{(0)}$.
       The growth rate of a $\vc{b}$-pure subgroup with
       strength $\tilde{n} \ge 1$ is estimated as
       $R^{(\tilde{n})} (P/2)$ with the perimeter $P$
       of the subgroup as in the classical paper
       by Burton et al.\cite{Burton:1951tr}
\end{enumerate}
Note that the critical distance
$2 \rho_c$ in the reduction procedure (i)
is close but not so close for a co-rotating pair.
In fact,
more accurate estimate of the growth rate by a co-rotating pair
$\vc{a}_1$ and $\vc{a}_2$ with $m_1 = m_2 = 1$ and
$|\vc{a}_1 - \vc{a}_2| = 2\rho_c$ is $R^{(2)} (2 \rho_c) = 2 R^{(0)} / (1 + 2 \omega_1 / \pi)
\approx (5/3) R^{(0)}$. 

We present some numerical results
verifying the above procedure.
The evolution equation of the following simulations is
\[
 V=6(1 - 0.05 \kappa),
\]
i.e., $\rho_c = 0.05$.
Note that $\tilde{d}_c \approx 0.474621$,
and $R^{(0)} \approx 6.320833$.
In the following simulations
the initial data is chosen as $u_0 (\vc{x}) = 0$.

We first
consider the following simple situation;
\renewcommand{\theenumi}{d\arabic{enumi}}
\begin{enumerate}
 \item \label{relay-1}
       $\vc{a}_1 = (-0.25,0; 2)$, $\vc{a}_2 = (0.25,0; 2)$
       (two independent centers with total strength $n=2$),
 \item \label{relay-2}
       $\vc{a}_1 = (-0.25,0; 2)$,
       $\vc{a}_2 = (0.25,0; 2)$,
       and $\vc{b} = (0,0; -1)$
       (a group of three centers with total strength $n=3$),
 \item \label{relay-3}
       $\vc{a}_1 = (-0.25,0; 2)$,
       $\vc{a}_2 = (0.25,0; 2)$,
       and $\vc{b} = (0,0; 1)$ 
       (a group of three centers with total strength $n=5$).
\end{enumerate}
\renewcommand{\theenumi}{\roman{enumi}}
The effective growth rates for these three cases are estimated as
\[
 \mbox{\eqref{relay-1}}:
 2 \times R^{(0)}
 \approx 12.641667, \quad
 \mbox{\eqref{relay-2}}:
 R^{(3)} (0.5)
 \approx 9.2343592, \quad
 \mbox{\eqref{relay-3}}:
 R^{(5)} (0.5)
 \approx 15.390599
\]
by the theory of Burton et al.\cite{Burton:1951tr}.
However, from our numerical results,  the computed $H(t)$
of  \eqref{relay-2} is almost identical to that of \eqref{relay-1},
even though three spirals seem to appear
from $\{ \vc{a}_1, \vc{a}_2, \vc{b} \}$ in the case \eqref{relay-2}.
See Figure \ref{growth: grouping_d1-3} and
Figure \ref{profile: grouping_d1-3}
for the details of $H(t)$ and the profiles of spirals,
respectively.
Note that $H(t)$ by \eqref{relay-3} is exactly
larger than those by \eqref{relay-2} or \eqref{relay-3}.
The numerical growth rates of \eqref{relay-1}--\eqref{relay-3}
computed the interval in $0.3 \le t \le 1.0$
are respectively:
\[
 \mbox{\eqref{relay-1}} \ R_\triangle \approx 12.846801, \quad
 \mbox{\eqref{relay-2}} \ R_\triangle \approx 12.841811, \quad
 \mbox{\eqref{relay-3}} \ R_\triangle \approx 14.445982.
 %
 %
\]
Summarizing the above discussion, we find that the cancellation phenomenon
is reflected to the number of spirals, but  to the effective growth rate.
\begin{figure}[htbp]
 \begin{center}
  \includegraphics[scale=1.0]{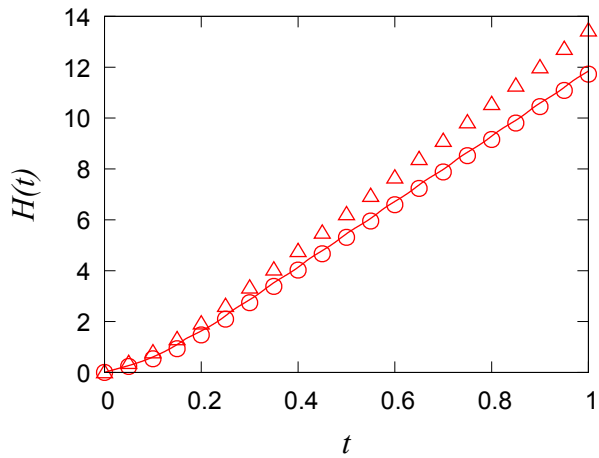}
  \caption{Graph of $H(t)$ by \eqref{relay-1}, \eqref{relay-2}
  and \eqref{relay-3}
  denoted by the solid line, circles and triangles,
  respectively.}
  \label{growth: grouping_d1-3}
 \end{center}
\end{figure}
\begin{figure}[htbp]
 \begin{center}
  \includegraphics[scale=1.0]{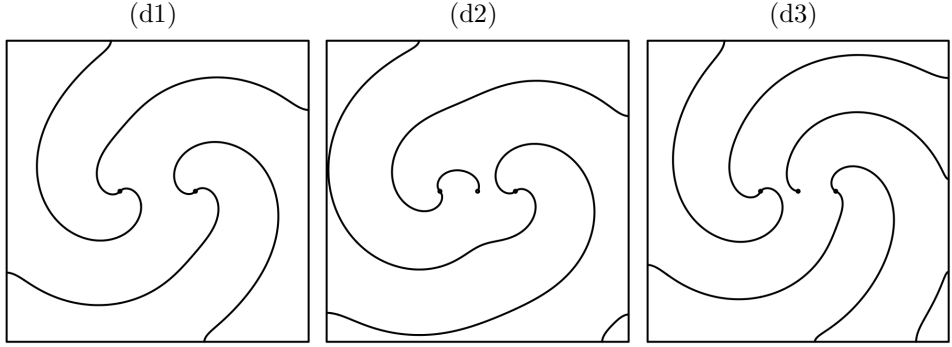}
  \caption{Profiles of spirals at $t=1$
  by \eqref{relay-1}, \eqref{relay-2}, and \eqref{relay-3}.}
  \label{profile: grouping_d1-3}
 \end{center}
\end{figure}

However,
if $\mathcal{A} = \{ \vc{a}_1, \ldots, \vc{a}_N \}$
make a co-rotating group
and there is a center $\vc{b}$ in the convex hull of $\mathcal{A}$
with the opposite rotational
orientation against to $\mathcal{A}$,
then the effective growth rate by $\mathcal{A} \cup \{ \vc{b} \}$
should be reduced from the rate by $\mathcal{A}$ only.
We present some numerical results
showing the above prediction.
Let us consider a co-rotating group of
4 centers 
$\mathcal{A} = \{ \vc{a}_1, \vc{a}_2, \vc{a}_3, \vc{a}_4 \}$
with
$\vc{a}_1 = (-0.21, 0.21;1)$, $\vc{a}_2 = (-0.21, -0.21; 1)$,
$\vc{a}_3 = (0.21, -0.21;1)$
and
\renewcommand{\theenumi}{e\arabic{enumi}}
\begin{enumerate}
 \item $\vc{a}_4 = (-0.07, - 0.07; 1)$,
       \label{convex hull-1}
 \item $\vc{a}_4 = (0.21, 0.21; 1)$.
       \label{convex hull-2}
\end{enumerate}
\renewcommand{\theenumi}{\roman{enumi}}
We examine the two situation
(a)just $\mathcal{A}$,
or (b)$\mathcal{A}$ with $\vc{b} = (0.07, 0.07; -1)$
for each situation.
See Figure \ref{location: convex hull}
for the location of $\mathcal{A}$
and $\vc{b}$,
in which the convex hull of $\mathcal{A}$
is denoted by the region enclosed by dotted line.
\begin{figure}[htbp]
 \begin{center}
  \includegraphics[scale=0.5]{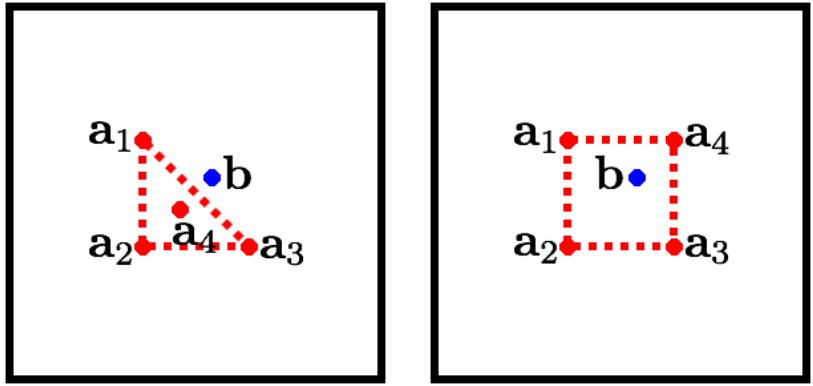}
  \caption{Location of centers for
  \eqref{convex hull-1}(left) and \eqref{convex hull-2}(right).
  Dotted line of each figure denotes the boundary of
  the convex hull of $\vc{a}_1, \vc{a}_2, \vc{a}_3$ and $\vc{a}_4$.}
  \label{location: convex hull}
 \end{center}
\end{figure}
For these situations
we calculate the numerical growth rate $R_\triangle$
with the data $H(t)$ for $0.3 \le t \le 1$,
and then we obtain
\[
 \begin{aligned}
  \mbox{\eqref{convex hull-1}}: \ &
  \mbox{(a)} \ R_\triangle \approx 9.328789, \quad 
  \mbox{(b)} \ R_\triangle \approx 9.154722, \\
  \mbox{\eqref{convex hull-2}}: \ &
  \mbox{(a)} \ R_\triangle \approx 8.132112, \quad 
  \mbox{(b)} \ R_\triangle \approx 7.290957.  
 \end{aligned}
\]
The reduction ratios of the growth rate
from the case (a) to (b)
in \eqref{convex hull-1} or \eqref{convex hull-2}
are about 2\% or 10\%, respectively.
The above results suggest that
the effective growth rate of a group of spirals is determined 
not only by the distance among the centers but also
whether a center with opposite rotational orientation is in 
the convex hull of the centers of the co-rotating group of spirals.


To clarify the effect of
the location of a center with the opposite
rotational orientation relative to a co-rotating group,
we examine the effective growth rates
of the configuration  $\mathcal{A} = \{ \vc{a}_1, \vc{a}_2, \vc{a}_3 \}$, where
\[
 \vc{a}_1 = (-0.21, 0.21; m), \quad
 \vc{a}_2 = (-0.21, -0.21; m), \quad
 \vc{a}_3 = (0.21, -0.21; m)
\]
and the center $\vc{b}$ with strength $-1$
at the following locations:
\renewcommand{\theenumi}{f\arabic{enumi}}
\begin{enumerate}
 \item \label{loc:f1}
       $\vc{b} = \vc{b}_1 = (-0.07, -0.07; -1)$
       (the center of mass of $\mathcal{A}$),
 \item \label{loc:f2}
       $\vc{b} = \vc{b}_2 = (-0.21, 0; -1)$,
 \item \label{loc:f3}
       $\vc{b} = \vc{b}_3 = (-0.27, 0; -1)$,
 \item \label{loc:f4}
       $\vc{b} = \vc{b}_4 = (-0.27, 0.21; -1)$,
 \item \label{loc:f5}
       $\vc{b} = \vc{b}_5 = (-0.27, -0.21; -1)$,
 \item \label{loc:f6}
       $\vc{b} = \vc{b}_6 = (-0.33, 0; -1)$,
 \item \label{loc:f7}
       $\vc{b} = \vc{b}_7 = (0, 0; -1)$,
 \item \label{loc:f8}
       $\vc{b} = \vc{b}_8 = (0.07, 0.07; -1)$.
\end{enumerate}
\renewcommand{\theenumi}{roman{enumi}}
See Figure \ref{location of centers}
for the location of $\mathcal{A}$ and $\vc{b}_j$.
Note that the situations $\vc{b}_2$ and $\vc{b}_7$ are different,
although they are on the boundary of the convex hull of $\mathcal{A}$.
In fact, $\vc{b}_2$ is between a co-rotating pair $\vc{a}_1$ and $\vc{a}_2$,
but $\vc{b}_7$ is between $\vc{a}_1$ and $\vc{a}_3$ which are not 
a co-rotating pair.
The locations $\vc{b}_3, \vc{b}_6, \vc{b}_8$ are just out of the convex hull.
These locations are for showing the recovery of
the effective growth rate to that of $\mathcal{A}$.
The locations $\vc{b}_4$ or $\vc{b}_5$ are for showing
the reduction of the growth rate
by the cancellation of strength between $\vc{b}_4$ and $\vc{a}_1$
or $\vc{b}_5$ and $\vc{a}_2$, respectively.
We also examine the growth rate
by $\{\vc{a}_1, \vc{a}_3 \}$, $\{ \vc{a}_2, \vc{a}_3 \}$ and
$\mathcal{A}$ as the benchmark tests.
\begin{figure}[htbp]
 \begin{center}
  \includegraphics[scale=0.3]{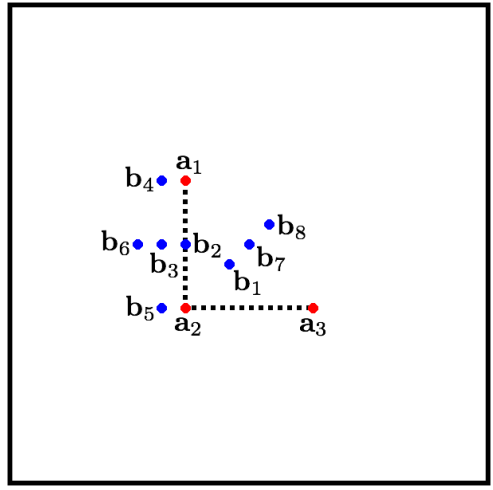}
  \caption{Location of each centers.
  The dotted line means the connection line
  of co-rotating group whose length are less than
  the effective distance.}
  \label{location of centers}
 \end{center}
\end{figure}

Table \ref{table: group 3-1}
lists the numerical growth rate with the data $H(t)$
for $0.3 \le t \le 1$.
\begin{table}[htbp]
 \begin{center}
  \caption{Growth rate by a group of
  3 co-rotating centers
  with $m=2$ and a center
  with the opposite rotation.} \label{table: group 3-1}
  \begin{tabular}[tb]{|c|c|}
   \hline
   Case & $R_\triangle$ \\
   \hline
   \eqref{loc:f1} & 13.268728 \\
   \eqref{loc:f2} & 13.383400 \\
   \eqref{loc:f3} & 13.580583 \\ 
   \eqref{loc:f4} & 13.283504 \\ 
   \eqref{loc:f5} & 12.811923 \\ 
   \eqref{loc:f6} & 13.833242 \\ 
   \eqref{loc:f7} & 13.553001 \\ 
   \eqref{loc:f8} & 13.808827 \\
   \hline
   $\{ a_1, a_3 \}$ & 12.724605 \\
   $\{ a_2, a_3 \}$ & 13.283220 \\
   $\mathcal{A}$ & 14.195001 \\
   \hline
  \end{tabular}
 \end{center}
\end{table}
In these simulations, the estimates of the growth rate
by Burton et al.\cite{Burton:1951tr} are as follows
\begin{align*}
 \mbox{by} \ \mathcal{A}: \
 & R^{(6)} (0.84) = \frac{6}{1 + 0.84 \omega_1 / (0.05\pi)} \times R^{(0)}
 \approx 13.692160, \\
 \mbox{by} \ \{ a_2, a_3 \}: \
 & R^{(4)} (0.42) = \frac{4}{1 + 0.42 \omega_1 / (0.05\pi)} \times R^{(0)}
 \approx 13.413502, \\
 \mbox{by} \ \{ a_1, a_3 \}: \
 & 2 \times R^{(0)} \approx 12.641667.
\end{align*}
Note that the growth rate by $\{ \vc{a}_1, \vc{a}_2 \}$ is the same
as that by $\{ \vc{a}_2, \vc{a}_3 \}$.
Also note that we use $L=P/2 = |\vc{a}_1 - \vc{a}_2| + |\vc{a}_2 - \vc{a}_3|$
with the perimeter $P$ of the group
for the estimate of the growth rate by $\mathcal{A}$,
which is not the perimeter of the convex hull of $\mathcal{A}$.
In fact, 
if we set $P = 0.84 + 0.42 \sqrt{2}$ as
the last one, 
then we obtain the estimate of the growth rate
as $R^{(6)} (0.5 \times P) = 15.105667$,
which is farther than $R^{(6)} (0.84)$
from $R_\triangle$ by $\mathcal{A}$.

According to \S 9.2 of Burton et al.\cite{Burton:1951tr},
the strength of the group by $\{ \vc{a}_1, \vc{a}_2, \vc{a}_3, \vc{b} \}$
in the case of (f1)--(f8)
decreases to 5 from 6, which is that of $\mathcal{A}$.
Then, the effective growth rate should be
$(5/6) \times R^{(6)} (0.84)$ or $5/6$ times of that by $\mathcal{A}$,
which is
\[
 14.195001 \times \frac{5}{6} = 11.8291675.
\]
However, the growth rates of the all cases of (f1)--(f8)
are larger than the above.
Our computation suggests that the reduction of the growth rate
is not caused by the cancellation of the strength.
On the other hand,
the growth rates by \eqref{loc:f4} or
\eqref{loc:f5} are close to those by
$\{ \vc{a}_2, \vc{a}_3 \}$ or $\{ \vc{a}_1, \vc{a}_3 \}$,
respectively.
These results should be caused by
the cancellation of the strength of $\vc{a}_1$
or $\vc{a}_2$,
thus the ($\vc{b}$-pure) pair $\{ \vc{a}_2, \vc{a}_3 \}$ or $\{ \vc{a}_1, \vc{a}_3 \}$
provides the maximum growth rate for each situation,
respectively.
The effective growth rates
by \eqref{loc:f1} and \eqref{loc:f3}
also seems to be caused by $\{ \vc{a}_2, \vc{a}_3 \}$.
The cases \eqref{loc:f3}, \eqref{loc:f6} and \eqref{loc:f8}
should recover the effective growth rate of $\mathcal{A}$
because $\vc{b}$ is far apart from the convex hull of $\mathcal{A}$.
See also Figure \ref{profile: grouping_f1-5}
for the profiles of spirals at $t=1$
for \eqref{loc:f1}, \eqref{loc:f4} and \eqref{loc:f5}.
For all cases one can find five spiral curves.
Note that the profile of the case \eqref{loc:f4}
makes a co-rotating pair from $\{ \vc{a}_2, \vc{a}_3 \}$ and
a single spiral from $\vc{a}_1$.
The profile of the case \eqref{loc:f5} is similar.
On the other hand,
the profile of the case \eqref{loc:f1}
shifts the three states of a co-rotating pair
and a single spiral.
\begin{figure}[htbp]
 \begin{center}
  \includegraphics[scale=1.0]{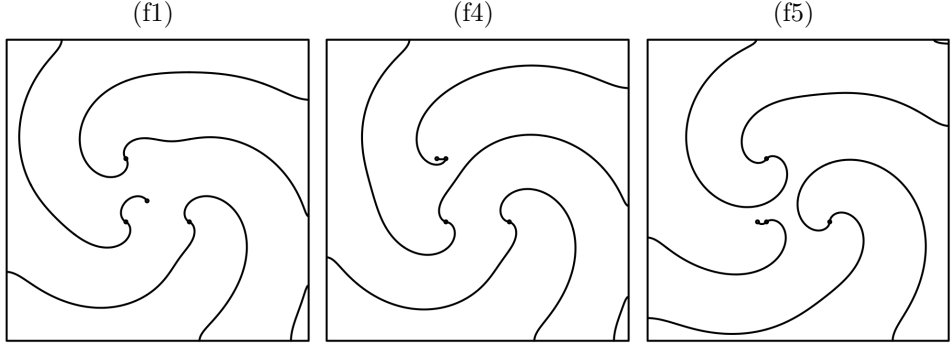}
  \caption{Profiles of spirals at $t=1$
  for \eqref{loc:f1}, \eqref{loc:f4}
  and \eqref{loc:f5}.}
  \label{profile: grouping_f1-5}
 \end{center}
\end{figure}

Finally, we conclude this section
with presenting the three results with $m=1$
and some remarks 
comparing the case between \eqref{loc:f2} and $\mathcal{A}$,
\eqref{loc:f4} and $\{ \vc{a}_2, \vc{a}_3\}$,
or \eqref{loc:f5} and $\{ \vc{a}_1, \vc{a}_3\}$,
respectively.
When we choose $m=1$ for the above cases
we obtain the following results.
\begin{align*}
 & \mbox{\eqref{loc:f2}}:
 R_\triangle \approx 7.378567,
 \quad
 \mathcal{A} : 
 R_\triangle \approx 7.123860, \\
 & \mbox{\eqref{loc:f4}}: R_\triangle \approx 6.658553,
 \quad
 \{ \vc{a}_2, \vc{a}_3 \}: 
 R_\triangle \approx 6.655576, \\
 & \mbox{\eqref{loc:f5}}: R_\triangle \approx 6.381851,
 \quad
 \{ \vc{a}_1, \vc{a}_3 \}: 
 R_\triangle \approx 6.381061.
\end{align*}
The case \eqref{loc:f2} of the above shows that
the growth rate of co-rotating group is not always reduced by
the center with the opposite rotation in the convex hull
of the group.
Such an acceleration may be caused by the effect
of the curvature as in the case of a pair
with the opposite rotations.
In fact, $|\vc{a}_1 - \vc{b}| = |\vc{a}_2 - \vc{b}| = 0.21 \approx 4 \rho_c$
in the case \eqref{loc:f2}.
On the other hand,  \eqref{loc:f4} or \eqref{loc:f5}
almost agree with that
by $\{ \vc{a}_2, \vc{a}_3 \}$ or $\{ \vc{a}_1, \vc{a}_3 \}$, respectively.
This means that the centers in an inactive pair
does not connect to other co-rotating centers.
Hence, we have case (i) in  the grouping procedure.
Finally, there is a case in which
the effective growth rate of $\mathcal{A} \cup \{ \vc{b} \}$
is greater than that of $\{ \vc{a}_2, \vc{a}_3 \}$.
In fact, if
$\vc{a}_1 = (-0.10,0.10; 5)$,
$\vc{a}_2 = (-0.10,-0.10; 5)$,
$\vc{a}_3 = (0.10,-0.10; 5)$,
and $\vc{b} = (0,0; -1)$,
then the numerical growth rates are
as follows.
\[
 \mathcal{A}: 
 R_\triangle \approx 49.707710, \quad
 \mathcal{A} \cup \{ \vc{b} \}: 
 R_\triangle \approx 47.063577, \quad
 \{ \vc{a}_2, \vc{a}_3 \}: 
 R_\triangle \approx 41.644654.
\]
Note that the estimate of the growth rate
with $\mathcal{A} \cup \{ \vc{b} \}$ by
the theory of Burton et al.
is $R^{(14)} (0.4) \approx 48.020802$,
and $14/15$ times of the growth rate by
$\mathcal{A}$ is also about $46.393862$.
}

\section{Conclusion}

In this paper, we study analytically and numerically the growth rate 
of a crystal surface growing by several screw dislocations.   
We carefully compare our estimates and simulation results with some of
the classical cases in the literature. We obtained new estimates on the growth rates 
for several different configurations (co-rotating pairs of spirals, spirals whose centers are co-linear, and groups of spirals), and we showed that these new rates were in agreement with the numerical simulations computed by the level set method proposed in the previous paper by the authors \cite{OTG:2015JSC}. 
We gave a new definition of the critical distance (of co-rotating pair)
with the view point of effective growth rate.
We also gave an improved estimate of the growth rate
by a co-rotating pair with an estimate of the
rotating single spiral by Ohara-Reid \cite{Ohara:1973ag}.
By arguments used in the above two items, we concluded that
the critical distance by Burton et al.\cite{Burton:1951tr} is too small.
We found that the growth rate by a pair of
opposite rotational screw dislocations, attains the maximum with
the distance between the spiral centers is
between $3.5 \rho_c$ and $4 \rho_c$.
We found that the distribution of screw dislocations on a line
influences to the growth rate.
We carefully studied how the growth rate depends on the 
distribution. 
For general group of spirals, we found that the growth rate
can be studied systematically
by the rates of the "effective" sub-groups of centers,
partitioned by the inter-distances.

\begin{acknowledgement}
 The authors are grateful to Professor Etsuro Yokoyama
 for valuable comments.
 The first author is partly supported by
 the Japan Society for the Promotion of Science through grants
 No. 26400158(Kiban C).
 The second author is partly supported by NSF grant DMS-1318975.
 The second author also thanks National Center
 for Theoretical Sciences (NCTS) of Taiwan
 for hosting his visits,
 during which part of this research was
 conducted.
 The third author is partly supported by the
 Japan Society for the Promotion of Science through
 grants No. 26220702(Kiban S),
 No. 17H01091(Kiban A),
 and No. 16H03948(Kiban B).
\end{acknowledgement}

\bibliographystyle{plain}
\bibliography{OTG}

\providecommand{\latin}[1]{#1}
\providecommand*\mcitethebibliography{\thebibliography}
\csname @ifundefined\endcsname{endmcitethebibliography}
  {\let\endmcitethebibliography\endthebibliography}{}
\begin{mcitethebibliography}{20}
\providecommand*\natexlab[1]{#1}
\providecommand*\mciteSetBstSublistMode[1]{}
\providecommand*\mciteSetBstMaxWidthForm[2]{}
\providecommand*\mciteBstWouldAddEndPuncttrue
  {\def\EndOfBibitem{\unskip.}}
\providecommand*\mciteBstWouldAddEndPunctfalse
  {\let\EndOfBibitem\relax}
\providecommand*\mciteSetBstMidEndSepPunct[3]{}
\providecommand*\mciteSetBstSublistLabelBeginEnd[3]{}
\providecommand*\EndOfBibitem{}
\mciteSetBstSublistMode{f}
\mciteSetBstMaxWidthForm{subitem}{(\alph{mcitesubitemcount})}
\mciteSetBstSublistLabelBeginEnd
  {\mcitemaxwidthsubitemform\space}
  {\relax}
  {\relax}

\bibitem[Osher and Sethian(1988)Osher, and
  Sethian]{Osher-Sethian-level-set:1988}
Osher,~S.; Sethian,~J.~A. Fronts propagating with curvature-dependent speed:
  algorithms based on {H}amilton-{J}acobi formulations. \emph{J. Comput. Phys.}
  \textbf{1988}, \emph{79}, 12--49\relax
\mciteBstWouldAddEndPuncttrue
\mciteSetBstMidEndSepPunct{\mcitedefaultmidpunct}
{\mcitedefaultendpunct}{\mcitedefaultseppunct}\relax
\EndOfBibitem
\bibitem[Sethian(1999)]{sethian-book:1999}
Sethian,~J.~A. \emph{Level set methods and fast marching methods}, 2nd ed.;
  Cambridge University Press: Cambridge, 1999; pp xx+378, Evolving interfaces
  in computational geometry, fluid mechanics, computer vision, and materials
  science\relax
\mciteBstWouldAddEndPuncttrue
\mciteSetBstMidEndSepPunct{\mcitedefaultmidpunct}
{\mcitedefaultendpunct}{\mcitedefaultseppunct}\relax
\EndOfBibitem
\bibitem[Osher and Fedkiw(2001)Osher, and Fedkiw]{Osher-Fedkiw:2000}
Osher,~S.; Fedkiw,~R.~P. Level set methods: an overview and some recent
  results. \emph{J. Comput. Phys.} \textbf{2001}, \emph{169}, 463--502\relax
\mciteBstWouldAddEndPuncttrue
\mciteSetBstMidEndSepPunct{\mcitedefaultmidpunct}
{\mcitedefaultendpunct}{\mcitedefaultseppunct}\relax
\EndOfBibitem
\bibitem[Giga(2006)]{Giga:2006}
Giga,~Y. \emph{Surface evolution equations: A level set approach}; Monographs
  in Mathematics; Birkh\"auser Verlag: Basel, 2006; Vol.~99; pp xii+264\relax
\mciteBstWouldAddEndPuncttrue
\mciteSetBstMidEndSepPunct{\mcitedefaultmidpunct}
{\mcitedefaultendpunct}{\mcitedefaultseppunct}\relax
\EndOfBibitem
\bibitem[Ohtsuka(2003)]{Ohtsuka:2003wi}
Ohtsuka,~T. {A level set method for spiral crystal growth}. \emph{Advances in
  Mathematical Sciences and Applications} \textbf{2003}, \emph{13},
  225--248\relax
\mciteBstWouldAddEndPuncttrue
\mciteSetBstMidEndSepPunct{\mcitedefaultmidpunct}
{\mcitedefaultendpunct}{\mcitedefaultseppunct}\relax
\EndOfBibitem
\bibitem[Ohtsuka \latin{et~al.}(2015)Ohtsuka, Tsai, and Giga]{OTG:2015JSC}
Ohtsuka,~T.; Tsai,~Y.-H.; Giga,~Y. A Level Set Approach Reflecting Sheet
  Structure with Single Auxiliary Function for Evolving Spirals on Crystal
  Surfaces. \emph{Journal of Scientific Computing} \textbf{2015}, \emph{62},
  831--874\relax
\mciteBstWouldAddEndPuncttrue
\mciteSetBstMidEndSepPunct{\mcitedefaultmidpunct}
{\mcitedefaultendpunct}{\mcitedefaultseppunct}\relax
\EndOfBibitem
\bibitem[Karma and Plapp(1998)Karma, and Plapp]{Karma:1998}
Karma,~A.; Plapp,~M. Spiral Surface Growth without Desorption. \emph{Phys. Rev.
  Lett.} \textbf{1998}, \emph{81}, 4444--4447\relax
\mciteBstWouldAddEndPuncttrue
\mciteSetBstMidEndSepPunct{\mcitedefaultmidpunct}
{\mcitedefaultendpunct}{\mcitedefaultseppunct}\relax
\EndOfBibitem
\bibitem[Kobayashi(2010)]{Kobayashi:1990th}
Kobayashi,~R. A brief introduction to phase field method. \emph{AIP Conf.
  Proc.} \textbf{2010}, \emph{1270}, 282--291\relax
\mciteBstWouldAddEndPuncttrue
\mciteSetBstMidEndSepPunct{\mcitedefaultmidpunct}
{\mcitedefaultendpunct}{\mcitedefaultseppunct}\relax
\EndOfBibitem
\bibitem[Miura and Kobayashi(2015)Miura, and Kobayashi]{Miura:2015CGD}
Miura,~H.; Kobayashi,~R. Phase-Field Modeling of Step Dynamics on Growing
  Crystal Surface: Direct Integration of Growth Units to Step Front.
  \emph{Crystal Growth \& Design} \textbf{2015}, \emph{15}, 2165--2175\relax
\mciteBstWouldAddEndPuncttrue
\mciteSetBstMidEndSepPunct{\mcitedefaultmidpunct}
{\mcitedefaultendpunct}{\mcitedefaultseppunct}\relax
\EndOfBibitem
\bibitem[Burton \latin{et~al.}(1951)Burton, Cabrera, and Frank]{Burton:1951tr}
Burton,~W.~K.; Cabrera,~N.; Frank,~F.~C. {The growth of crystals and the
  equilibrium structure of their surfaces}. \emph{Philosophical Transactions of
  the Royal Society of London. Series A. Mathematical and Physical Sciences}
  \textbf{1951}, \emph{243}, 299--358\relax
\mciteBstWouldAddEndPuncttrue
\mciteSetBstMidEndSepPunct{\mcitedefaultmidpunct}
{\mcitedefaultendpunct}{\mcitedefaultseppunct}\relax
\EndOfBibitem
\bibitem[Not()]{Note-1}
Ref.~\citenum {OTG:2015JSC}, Definition 3,5\relax
\mciteBstWouldAddEndPuncttrue
\mciteSetBstMidEndSepPunct{\mcitedefaultmidpunct}
{\mcitedefaultendpunct}{\mcitedefaultseppunct}\relax
\EndOfBibitem
\bibitem[Goto \latin{et~al.}(2008)Goto, Nakagawa, and Ohtsuka]{Goto:2008hy}
Goto,~S.; Nakagawa,~M.; Ohtsuka,~T. {Uniqueness and existence of generalized
  motion for spiral crystal growth}. \emph{Indiana University Mathematics
  Journal} \textbf{2008}, \emph{57}, 2571--2599\relax
\mciteBstWouldAddEndPuncttrue
\mciteSetBstMidEndSepPunct{\mcitedefaultmidpunct}
{\mcitedefaultendpunct}{\mcitedefaultseppunct}\relax
\EndOfBibitem
\bibitem[Not()]{Note-2}
Ref.~\citenum {OTG:2015JSC}, \S 3.1\relax
\mciteBstWouldAddEndPuncttrue
\mciteSetBstMidEndSepPunct{\mcitedefaultmidpunct}
{\mcitedefaultendpunct}{\mcitedefaultseppunct}\relax
\EndOfBibitem
\bibitem[Not()]{Note-3}
Ref.~\citenum {OTG:2015JSC}, \S 3.1\relax
\mciteBstWouldAddEndPuncttrue
\mciteSetBstMidEndSepPunct{\mcitedefaultmidpunct}
{\mcitedefaultendpunct}{\mcitedefaultseppunct}\relax
\EndOfBibitem
\bibitem[Cabrera and Levine(1956)Cabrera, and Levine]{Cabrera-Levine:1956}
Cabrera,~N.; Levine,~M.~M. XLV. On the dislocation theory of evaporation of
  crystals. \emph{Philosophical Magazine} \textbf{1956}, \emph{1},
  450--458\relax
\mciteBstWouldAddEndPuncttrue
\mciteSetBstMidEndSepPunct{\mcitedefaultmidpunct}
{\mcitedefaultendpunct}{\mcitedefaultseppunct}\relax
\EndOfBibitem
\bibitem[Ohara and Reid(1973)Ohara, and Reid]{Ohara:1973ag}
Ohara,~M.; Reid,~R.~C. \emph{Modeling Crystal growth rates from solution};
  Prentice-Hall Inc., 1973\relax
\mciteBstWouldAddEndPuncttrue
\mciteSetBstMidEndSepPunct{\mcitedefaultmidpunct}
{\mcitedefaultendpunct}{\mcitedefaultseppunct}\relax
\EndOfBibitem
\bibitem[Ohtsuka(2014)]{Ohtsuka:2014suriken}
Ohtsuka,~T. Evolution of crystal surface by a single screw dislocation with
  multiple spiral steps. \emph{S\=urikaisekikenky\=usho K\=oky\=uroku}
  \textbf{2014}, 11--20, Mathematical analysis of pattern formation arising in
  nonlinear phenomena (Kyoto, 2013)\relax
\mciteBstWouldAddEndPuncttrue
\mciteSetBstMidEndSepPunct{\mcitedefaultmidpunct}
{\mcitedefaultendpunct}{\mcitedefaultseppunct}\relax
\EndOfBibitem
\bibitem[Ohtsuka(2010)]{Ohtsuka:OWR2010}
Ohtsuka,~T. Level set method for spiral crystal growth and surface evolution.
  \emph{Oberwolfach report} \textbf{2010}, \emph{7}, 291--294\relax
\mciteBstWouldAddEndPuncttrue
\mciteSetBstMidEndSepPunct{\mcitedefaultmidpunct}
{\mcitedefaultendpunct}{\mcitedefaultseppunct}\relax
\EndOfBibitem
\bibitem[Not()]{Note-4}
Ref.~\citenum {Burton:1951tr}, \S 9.2\relax
\mciteBstWouldAddEndPuncttrue
\mciteSetBstMidEndSepPunct{\mcitedefaultmidpunct}
{\mcitedefaultendpunct}{\mcitedefaultseppunct}\relax
\EndOfBibitem
\end{mcitethebibliography}

\end{document}